\documentclass[times,tighten,twocolumn]{aastex631}
\usepackage{amsmath}	% Advanced maths commands
\usepackage{amssymb}	% Extra maths symbols
\usepackage{acronym} % Acronyms
\usepackage{longtable}
\usepackage{multirow}
\usepackage{units}

\submitjournal{\apj}
\shorttitle{Multiband observations of GLEAM-X\,J162759.5-523504.3}
\shortauthors{Rea et al.}
\begin{document}

\title{Constraining the nature of the 18\,min periodic radio transient GLEAM-X\,J162759.5-523504.3 via multi-wavelength observations and magneto-thermal simulations}

%% Authors in sparse order
% From ICE-CSIC: theoretical models and X-ray observations: C. Dehman, M. Ronchi, V. Graber, D. Vigano', A. Borghese, F. Coti Zelati, N. Rea
% From the discovery paper and Chandra proposal: N. Hurley-Walker
% From the optical observations: D. de Martino, D. Buckely, J. Brink
% Collaborations: C. Kouveliotou, E. Gogus

\correspondingauthor{Nanda Rea}
 \email{rea@ice.csic.es}
 
\author[0000-0003-2177-6388]{N. Rea}
\affiliation{Institute of Space Sciences (ICE), CSIC, Campus UAB, Carrer de Can Magrans s/n, E-08193, Barcelona, Spain}
\affiliation{Institut d'Estudis Espacials de Catalunya (IEEC), Carrer Gran Capit\`a 2--4, E-08034 Barcelona, Spain} 

\author[0000-0001-7611-1581]{F. Coti Zelati}
\affiliation{Institute of Space Sciences (ICE), CSIC, Campus UAB, Carrer de Can Magrans s/n, E-08193, Barcelona, Spain}
\affiliation{Institut d'Estudis Espacials de Catalunya (IEEC), Carrer Gran Capit\`a 2--4, E-08034 Barcelona, Spain}

\author[0000-0003-0554-7286]{C. Dehman}
\affiliation{Institute of Space Sciences (ICE), CSIC, Campus UAB, Carrer de Can Magrans s/n, E-08193, Barcelona, Spain}
\affiliation{Institut d'Estudis Espacials de Catalunya (IEEC), Carrer Gran Capit\`a 2--4, E-08034 Barcelona, Spain}

\author[0000-0002-5119-4808]{N. Hurley-Walker}
\affiliation{International Centre for Radio Astronomy Research, Curtin University, Kent St, Bentley WA 6102, Australia}

\author[0000-0002-5069-4202]{D. De Martino}
\affiliation{INAF--Osservatorio Astronomico di Capodimonte, Salita Moiariello 16, I-80131 Napoli, Italy}

\author[0000-0003-2506-6041]{A. Bahramian}
\affiliation{International Centre for Radio Astronomy Research, Curtin University, Kent St, Bentley WA 6102, Australia}

\author[0000-0002-7004-9956]{D. A. H. Buckley}
\affiliation{South African Astronomical Observatory, PO Box 9, Observatory Road, Observatory 7935, Cape Town, South Africa}
\affiliation{Department of Astronomy, University of Cape Town, Private Bag X3, Rondebosch 7701, South Africa}

\author[0000-0003-0030-7566]{J. Brink}
\affiliation{South African Astronomical Observatory, PO Box 9, Observatory Road, Observatory 7935, Cape Town, South Africa}
\affiliation{Department of Astronomy, University of Cape Town, Private Bag X3, Rondebosch 7701, South Africa}

\author[0000-0002-4485-6471]{A. Kawka}
\affiliation{International Centre for Radio Astronomy Research, Curtin University, Kent St, Bentley WA 6102, Australia}

\author[0000-0003-1018-8126]{J. A. Pons}
\affiliation{Departament de F\'isica Aplicada, Universitat d’Alacant, 03690 Alicante, Spain}

\author[0000-0001-7795-6850]{D. Vigan\`o}
\affiliation{Institute of Space Sciences (ICE), CSIC, Campus UAB, Carrer de Can Magrans s/n, E-08193, Barcelona, Spain}
\affiliation{Institut d'Estudis Espacials de Catalunya (IEEC), Carrer Gran Capit\`a 2--4, E-08034 Barcelona, Spain} 

\author[0000-0002-6558-1681]{V. Graber}
\affiliation{Institute of Space Sciences (ICE), CSIC, Campus UAB, Carrer de Can Magrans s/n, E-08193, Barcelona, Spain}
\affiliation{Institut d'Estudis Espacials de Catalunya (IEEC), Carrer Gran Capit\`a 2--4, E-08034 Barcelona, Spain} 

\author[0000-0003-2781-9107]{M. Ronchi}
\affiliation{Institute of Space Sciences (ICE), CSIC, Campus UAB, Carrer de Can Magrans s/n, E-08193, Barcelona, Spain}
\affiliation{Institut d'Estudis Espacials de Catalunya (IEEC), Carrer Gran Capit\`a 2--4, E-08034 Barcelona, Spain} 

\author{C. Pardo Araujo}
\affiliation{Institute of Space Sciences (ICE), CSIC, Campus UAB, Carrer de Can Magrans s/n, E-08193, Barcelona, Spain}
\affiliation{Institut d'Estudis Espacials de Catalunya (IEEC), Carrer Gran Capit\`a 2--4, E-08034 Barcelona, Spain} 

\author[0000-0001-8785-5922]{A. Borghese}
\affiliation{Institute of Space Sciences (ICE), CSIC, Campus UAB, Carrer de Can Magrans s/n, E-08193, Barcelona, Spain}
\affiliation{Institut d'Estudis Espacials de Catalunya (IEEC), Carrer Gran Capit\`a 2--4, E-08034 Barcelona, Spain} 

\author[0000-0002-0430-6504]{E. Parent}
\affiliation{Institute of Space Sciences (ICE), CSIC, Campus UAB, Carrer de Can Magrans s/n, E-08193, Barcelona, Spain}
\affiliation{Institut d'Estudis Espacials de Catalunya (IEEC), Carrer Gran Capit\`a 2--4, E-08034 Barcelona, Spain} 

\def\xmm {\emph{XMM--Newton}}
\def\cxo {\emph{Chandra}}
\def\nustar {\emph{NuSTAR}}
\def\rst {\emph{ROSAT}}
\def\swift {\emph{Swift}}
\def\nicer {\emph{NICER}}
\def\hxmt {\emph{Insight}-HXMT}
\def\pks {Parkes}

\def\flux {\mbox{erg\,cm$^{-2}$\,s$^{-1}$}}
\def\lum {\mbox{erg\,s$^{-1}$}}
\def\nh {N$_{\rm H}$}
\def\kms  {\rm \ km \, s^{-1}}
\def\cms  {\rm \ cm \, s^{-1}}
\def\gs   {\rm \ g  \, s^{-1}}
\def\cmtre {\rm \ cm^{-3}}
\def\cm2 {\rm \ cm$^{-2}$}
\def\ss {\mbox{s\,s$^{-1}$}}
\def\chisq {$\chi ^{2}$}
\def\rchisq {$\chi_{r} ^{2}$}

\def\arc{\mbox{$^{\prime\prime}$}}
\def\arcmin{\mbox{$^{\prime}$}}
\def\deg{\mbox{$^{\circ}$}}

\def\rsun {~R_{\odot}}
\def\msun {~M_{\odot}}
\def\mdotav {\langle \dot {M}\rangle }

\def\gleamfirst{\mbox{GLEAM-X\,J162759.5-523504.3}}
\def\gleam {\mbox{GLEAM-X\,J1627}}
\def\mtp{\mbox{MTP0013}}
\def\rcw{\mbox{1E\,161348-5055}}
\def\lowbsgr{\mbox{SGR\,0418+5729}}

\def\uu {4U\,0142$+$614}
\def\ee {1E\,1048.1$-$5937}
\def\kes {1E\,1841$-$045}
\def\aa {1E\,1547$-$5408}
\def\axj {AX\,J1844$-$0258}
\def\rxs {1RXS\,J1708$-$4009}
\def\xte{XTE\,J1810$-$197}
\def\smc{CXOU\,J0100$-$7211\,}
\def\wes{CXOU\,J1647$-$4552}
\def\ea {1E\,2259$+$586}
\def\ctb{CXOU\,J171405.7$-$381031}
\def\sgra{SGR\,1806$-$20}
\def\sgrb{SGR\,1900$+$14}
\def\sgrd{SGR\,1627$-$41}
\def\sgre{SGR\,0501$+$4516}
\def\sgrf{SGR\,1935+2154}
\def\lowba{SGR\,0418$+$5729}
\def\sgrg{SGR\,1833$-$0832}
\def\lowbb{Swift\,J1822.3$-$1606}
\def\galmag{PSR\,J1745$-$2900}
\def\sgras{Sgr\,A$^{\star}$}
\def\sgrh{SGR\,1801$-$21}
\def\sgri{SGR\,2013$+$34}
\def\psr{PSR\,1622$-$4950}
\def\hbpsr{PSR\,J1846$-$0258}
\def\radiohb{PSR\,J1119$-$6127}
\def\coronamag{Swift\,J1818.0$-$1607}
\def\sgrl{SGR\,J1830$-$0645}

%%%%%%%%%%%%%%%%%%%%%%%%%%%%%%%%%%%%%%%%%%%%%%%%%%%%%%%%%%%%%%%%

\begin{abstract}

We observed the periodic radio transient \gleamfirst\ (\gleam) using the {\em Chandra X-ray Observatory} for about $30$\,ks on January 22--23, 2022, simultaneously with radio observations from MWA, MeerKAT and ATCA. Its radio emission and 18\,min periodicity led the source to be tentatively interpreted as an extreme magnetar or a peculiar highly magnetic white dwarf.
The source was not detected in the 0.3--8\,keV energy range with a $3\sigma$ upper-limit on the count rate of $3\times10^{-4}$\,counts\,s$^{-1}$. No radio emission was detected during our X-ray observations either. Furthermore, we studied the field around \gleam\ using archival ESO and DECam data, as well as recent SALT observations. Many sources are present close to the position of \gleam, but only two within the 2$\arcsec$ radio position uncertainty.
Depending on the assumed spectral distribution, the upper limits converted to an X-ray luminosity of $L_{X}<6.5\times10^{29}$\,\lum\ for a blackbody with temperature $kT=0.3$\,keV, or $L_{X}<9\times10^{29}$\,\lum\ for a power-law with photon index $\Gamma = 2$ (assuming a 1.3\,kpc distance). Furthermore, we performed magneto-thermal simulations for neutron stars considering crust- and core-dominated field configurations. Based on our multi-band limits, we conclude that: i) in the magnetar scenario, the X-ray upper limits suggest that \gleam\ should be older than $\sim$1 Myr, unless it has a core-dominated magnetic field or has experienced fast-cooling; ii) in the white dwarf scenario, we can rule out most binary systems, a hot sub-dwarf and a hot magnetic isolated white dwarf ($T\gtrsim10.000$\,K), while a cold isolated white dwarf is still compatible with our limits.

\end{abstract}

%%%%%%%%%%%%%%%%%%%%%%%%%%%%%%%%%%%%%%%%%%%%%%%%%%%%%%
\section{Introduction} 
\label{sec:intro}

%------------------------------------------------------------------

\begin{table*}[t]
\centering 
\caption{X-ray and Radio observations.} 
\label{tab:radio_obslog}
\begin{tabular}{@{}clrccc}
	\hline
	\hline
Instrument  		&  Start Time (UTC) 			& End Time (UTC) 		& Exposure$^{a}$ 	& Obs.\,ID  		& Mode$^{b}$\\
 			        & \multicolumn{2}{c}{(YYYY Mmm DD hh:mm:ss)}		    & (ks) 				&    &   \\ \hline
\cxo                &  2022 Jan 22 20:50:23 		& 2022 Jan 23 03:04:34	& 19.82 	        & 26228			& TE VF (3.041\,s) \\
\cxo 			    & 2022 Jan 23 06:03:38 		    & 2022 Jan 23 09:25:25	& 9.98 	            & 26282			& TE VF (3.141\,s)\\
\hline
\hline
Instrument  		& Start Time (UTC) 			& End Time (UTC) 		& Integration time$^{c}$ 	& Observing Band & 3$\sigma$ limit \\
 				    & \multicolumn{2}{c}{(YYYY Mmm DD hh:mm:ss)}		& (minutes) 				& (GHz)  & ($\mu$Jy\,beam$^{-1}$) \\ \hline
MWA		& 2022 Jan 21 23:45:58		& 2022 Jan 22 01:24:38	        & 111 			& 0.17 -- 0.2 & -- \\
MWA		& 2022 Jan 22 23:45:58 		& 2022 Jan 23 01:24:38	        & 111 			& 0.17 -- 0.2 & 5000 \\
MeerKAT & 2022 Jan 23 03:32:49      & 2022 Jan 23 05:12:48          & 70            & 0.58 -- 1.1 &  84 \\
ATCA	& 2022 Jan 22 15:00:18		& 2022 Jan 22 19:41:42          & 252 		    & 4 -- 6 & 72 \\
ATCA	& 2022 Jan 22 15:00:18	    & 2022 Jan 22 19:41:42	        & 252 		    & 8 -- 10 & 60 \\
\hline
\hline
\end{tabular}
\begin{list}{}{}
\item[$^{a}$] Deadtime corrected on-source time.
\item[$^{b}$] TE: Timed Exposure, VF: Very Faint telemetry format; the temporal resolution is given in parentheses. 
\item[$^{c}$] On-source only; not including time spent on calibrators.
\end{list}

\end{table*}
%------------------------------------------------------------------

\begin{figure*}
 \includegraphics[width=0.99\textwidth]{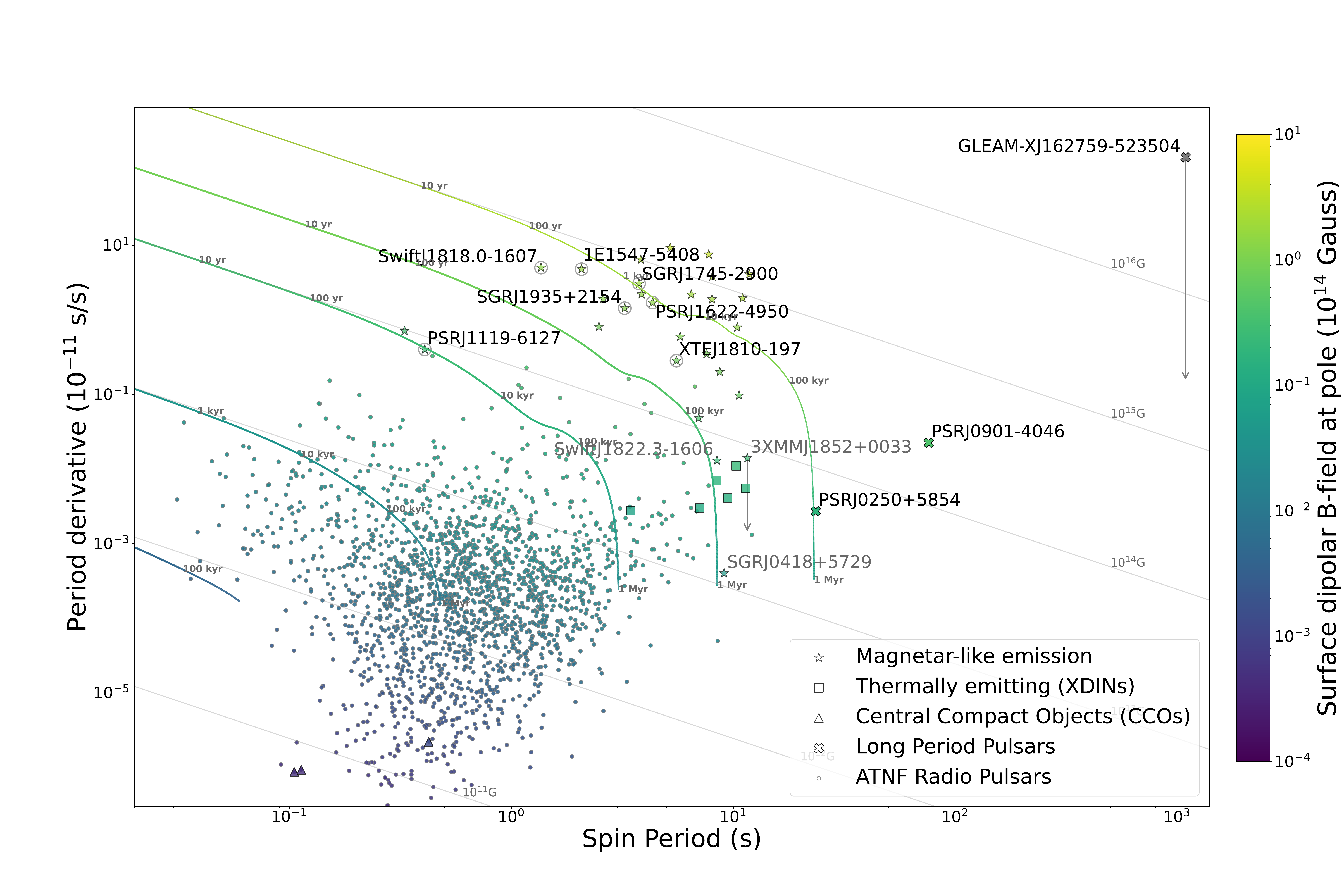}
\vspace{-0.4cm}
\caption{$P$-$\dot{P}$ diagram for different neutron star classes. The names are shown for radio-magnetars (also displayed with a gray circle), low-field magnetars, and long-period pulsars (including \gleam). Light gray lines correspond to constant magnetic fields, while the colored curves assumed magnetic field decay in the crust (see \S\ref{sec:discussion} for details). Colors refer to the surface dipolar magnetic field strength at the pole. }
\label{fig:P_Pdot_B}
\end{figure*}

The spin-period ($P$) evolution of a pulsar is driven by the combination of several factors: the presence of accretion during its lifetime either from supernova fall-back or a companion star, the dissipation of the magnetic field over time due to currents inside and outside the neutron star crust, the consumption of rotational energy via dipolar spin-down emission, occasional glitch events, etc. \citep{Lyne1985,Manchester1977}. The population of rotational-powered radio pulsars has been observed to have spin periods in the range $P\sim0.0014-12$\,s \citep{ATNFcatalog}. The majority of the observed population clusters around $P\sim1$\,s, with the extremes being populated on the fast side by rapidly spinning recycled millisecond pulsars and on the slow side by magnetars. The spin distribution of classical magnetars \citep{Kaspi2017,Esposito2021} ranges between $\sim$1.4\,s for \coronamag, a young radio magnetar recently discovered during an outburst \citep{Esposito2020}, and $\sim$12\,s for the bright hard X-ray emitting magnetar \kes\  \citep{vasisht97}. However, the discovery of magnetar-like activity from the high magnetic field, rotation-powered pulsars \radiohb\ \citep[$P\sim0.11$\,s;][]{gogus16} and \hbpsr\ \citep[$P\sim0.3$\,s;][]{gavriil08}, as well as from \rcw\ at the center of the supernova remnant RCW103 \citep[$P\sim6.67$\,hr;][]{Rea2016,D'Ai2016}, has enlarged the historical magnetar spin-period range. Furthermore, a few rotational-powered radio pulsars have been recently discovered with periods larger than the classical magnetar range: PSR\,J1903$+$0433 \citep[$P\sim14.1$\,s;][]{Han2021}, PSR\,J0250$+$5854 \citep[$P\sim23.5$\,s;][]{Tan2018}
and PSR\,J0901$-$4046 \citep[$P\sim76$\,s;][]{Caleb2022} (see also Fig.\,\ref{fig:P_Pdot_B}).

On the other hand, the spin periods of magnetic white dwarfs range between $\sim$0.019--10$^{4}$\,hr \citep{Brinkworth2013,Ferrario2020,Kilic2021}. Magnetic white dwarfs have been observed both isolated (about 600 detected thus far) and in interacting binaries (about 200 detected), with magnetic fields reaching $\sim$10$^{9}$\,G. Periodic radio and optical emission has only been detected from the binary star AR\,Sco at the beat frequency between its spin and orbital periods \citep{marsh16}. The incoherent nature of the pulsed radio emission from AR\,Sco has pointed towards models involving particle acceleration due to the interaction between the two stars rather than canonical pulsar radio emission \citep{geng2016}.

A peculiar radio transient with a periodicity of $\sim$1091\,s (\gleamfirst{}; \gleam\ hereafter) was discovered in archival data taken by the Murchison Widefield Array \citep[MWA;][]{2013PASA...30....7T,2018PASA...35...33W}. This source was found to be active during January--March 2018 with 5--40\,Jy bright radio pulses lasting about 10--30\,s and repeating with an 18-min periodicity \citep{Hurley-Walker2022}. The bright pulses have a linear polarization degree of 88$\pm$1\% and a dispersion measure DM = 57$\pm$1\,pc\,cm$^{-3}$, the latter resulting in a distance of $1.3\pm0.5$\,kpc according to the Galactic electron-density model by \cite{Yao2017}.
The observed radio luminosity and emission timescales pointed to a coherent emission process. On the one hand, the long spin period would be explained by a rotating magnetic white dwarf emitting as a pulsar-like dipole. On the other hand, the source radio properties resembled those of radio magnetars. In any of the above-mentioned cases, the nature of this source would be quite extreme within these source classes \citep{Hurley-Walker2022}.
In this work we report on X-ray observations of \gleam\ performed with the {\em Chandra X-ray Observatory} on January 22-23, 2022 (\S\ref{sec:X_obs}) and simultaneous radio observations using MWA, MeerKAT, and the Australia Telescope Compact Array (ATCA; \S\ref{sec:radio_obs}). 
We also report on optical and near infrared (nIR) data sets collected in ESO public survey projects and in the DECam Plane Survey, as well as optical data acquired in recent SALT observations (\S\ref{sec:opt_obs}). 
We use these observations to constrain the nature of this periodic radio transient both in the neutron star and white dwarf scenarios. We report the results in \S\ref{sec:results} and discuss them in \S\ref{sec:discussion}. Conclusions follow in \S\ref{sec:conclusions}.

%
%----------------------------------------------------
%----------------------------------------------------

%%%%%%%%%%%%%%%%%%%%%%%%%%%%%%%%%%%%%%%%%%%%%%%%%%%%%%%%%%%%%%%%
\section{Observations and data analysis} 

\subsection{X-ray observations: Chandra}
\label{sec:X_obs}

The {\em Chandra X-ray Observatory} observed \gleam\ twice using the Advanced CCD Imaging Spectrometer (ACIS; \citealt{garmire03}) instrument, from 2022 January 22 at 20:51:32 to January 23 at 03:05:43 TT (ObsID: 26228), and then again on 2022 January 23 from 06:04:47 to 09:26:34 TT (ObsID: 26282). All observations were performed in timed exposure imaging mode with VERY FAINT (VF) telemetry format. The source was positioned on the back-illuminated ACIS-S3 CCD at the nominal target position (R.A. = 16$^\mathrm{h}$27$^\mathrm{m}$59$\fs$5, decl. = --52$^{\circ}$35$^{\prime}$04$\farcs$3; J2000.0; uncertainty of 2\arc; \citealt{Hurley-Walker2022}). 

Standard processing of the data was performed by the Chandra X-ray Center to Level 1 and Level 2 (processing software DS 10.10.2.1). The data were reprocessed using the \textsc{ciao} software (version 4.14; \textsc{caldb} 4.9.6). We used the latest ACIS gain map, and applied the time-dependent gain and charge transfer inefficiency corrections. The data were then filtered for bad event grades and only good time intervals were used. No high background events were detected, resulting in an exposure time of about 20.1 and 10.1\,ks for the first and second observation, respectively.

We did not detect any significant X-ray emission at the radio position of \gleam\ when co-adding the two \cxo\ observations (total on-source livetime of 29.8\,ks).  
Specifically, we detected only one photon in the 0.3--8\,keV energy range within a 2\arc\ error circle centered on the source position. Taking a 3$\sigma$ upper limit of 8.9 photons \citep{gehrels86}, we can infer an upper limit on the X-ray quiescent count rate of \gleam\ of $\sim2.9\times10^{-4}$ counts\,s$^{-1}$ (at 3$\sigma$ c.l.). 
We checked this number using different approaches. We extracted source counts from circular regions with a radius of 0.8\arc\ (enclosing 90\% of the point spread function region at 1\,keV), 2\arc\ or 3\arc. For the background, we also assumed different extraction regions: an annulus with an inner radius equal to the source radius and an outer radius 5 times larger, or circles of the same size as the source extraction regions, far from the aim point but located on the same CCD. We used the \textsc{ciao} \textsc{srcflux} tool to extract the upper limits of the source count-rate using all the different combinations of source and background extractions. In all cases, the tool gives a 3$\sigma$ limit on the background-subtracted count rate at the position of the radio source of $3\times10^{-4}$\,counts\,s$^{-1}$ (0.3--8\,keV) after merging the event files from the two observations.

%----------------------------------------------------
\begin{figure*}
\centering
\vspace{0.3cm}
\includegraphics[width =0.7\textwidth]{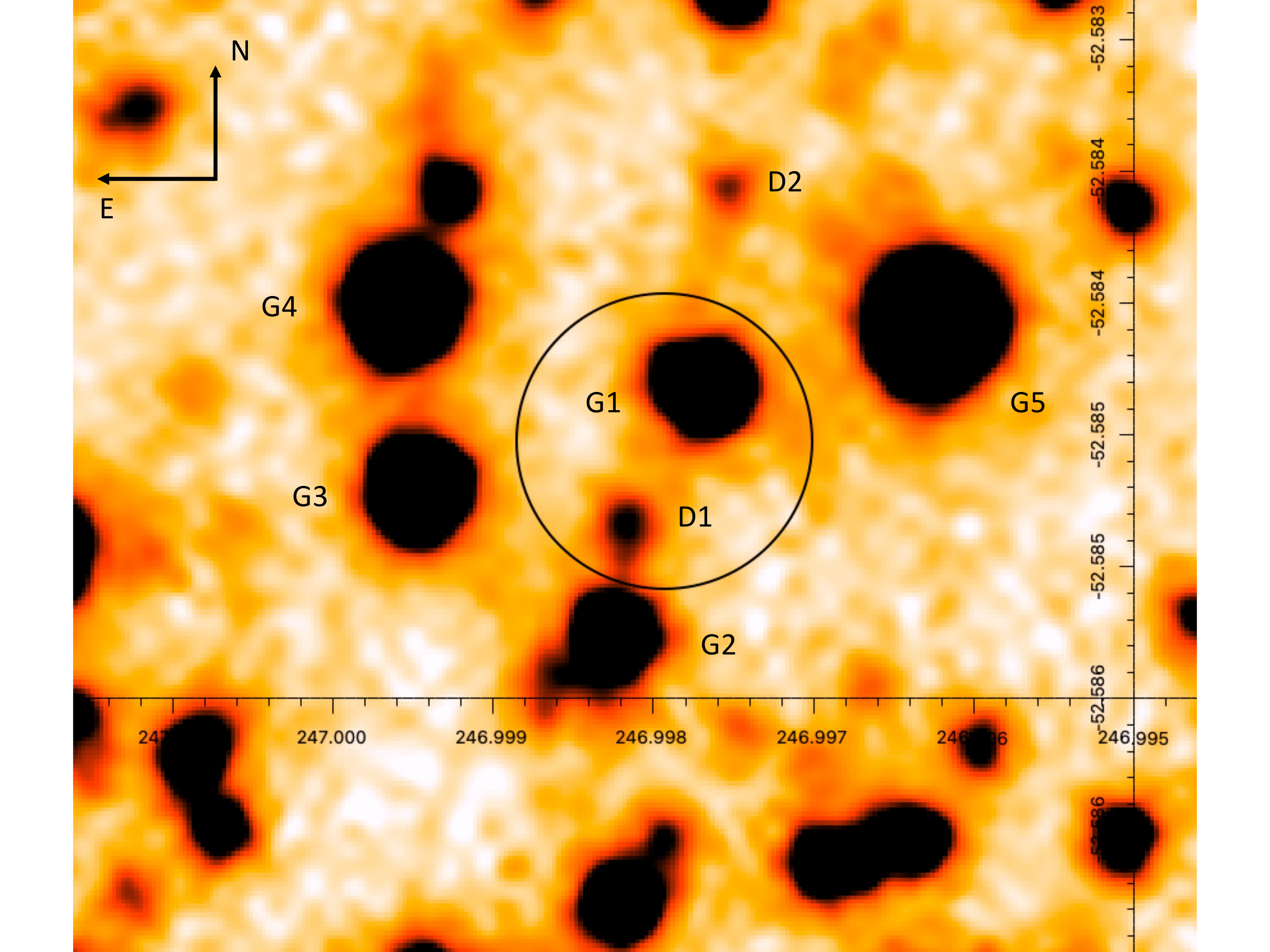}
\caption{The $z$-band DECAPS image of the field of \gleam. The position of the radio source and its 1-$\sigma$ error circle (with a radius of 2$\arc$) are reported. The positions of the five {\em Gaia} (G1, G2, G3, G4, G5) and two DECAPS (D1, D2) objects within 4$\arc$ are also reported. North is up and East is left.}
\label{fig:decaps_FoV}
\end{figure*}  
%----------------------------------------------------

%%%%%%%%%%%%%%%%%%%%%%%%%%%%%%%%%%%%%%%%%%%%%%%%%%%%%%%%%%%%%%%%
\subsection{Radio observations: MWA, MeerKAT and ATCA}
\label{sec:radio_obs}

We performed near-contemporaneous radio observations with three different radio telescopes: the MWA, MeerKAT, and ATCA, spanning a frequency range of 170\,MHz--9\,GHz. We found no pulsed or continuum radio emission at the location of \gleam. The observation properties and the derived (3$\sigma$) upper limits of the flux density are shown in Tab.\,\ref{tab:radio_obslog}.

\subsubsection{MWA observations}

We observed with the Phase~\textsc{II} extended configuration of the MWA using 5-minute pointed snapshots at 170--200\,MHz as \gleam\ transited (i.e. when the primary beam sensitivity was highest). 104~tiles were functional at the time of observing. We reduced the data using the download, calibration, and imaging stages of the GLEAM-X pipeline\footnote{\url{https://github.com/tjgalvin/GLEAM-X-pipeline}} \citep[][in press]{2022arXiv220412762H}.
Stokes~I imaging was performed with WS\textsc{clean} \citep{2014MNRAS.444..606O} using standard GLEAM-X settings, resulting in a restoring beam size of $1'$. Primary beam correction was performed using the most up-to-date MWA beam model \citep{2017PASA...34...62S}. At these frequencies, the source dispersion measure of 57\,pc\,cm$^{-3}$ causes $\sim2$\,s of smearing across the band, so there is no need to perform dedispersion to make a detection. Since the pulse profile of \gleam\ was previously 30--60\,s wide, and the integration time is 4\,s, we split the data into 32-s intervals and folded at $P=1091$\,s. No pulsed emission was observed to a 3$\sigma$ limit of 16\,mJy\,beam$^{-1}$ per (32\,s-long) phase bin. No bright single pulses were observed in any individual timestep, down to a 3$\sigma$ upper limit of 100\,mJy\,beam$^{-1}$. The RMS noise $\sigma$ in each 5-min snapshot is 10\,mJy\,beam$^{-1}$. By stacking all 222~min of observing time, we obtained a 3$\sigma$ upper limit in the mosaic of 5\,mJy\,beam$^{-1}$.

\subsubsection{MeerKAT observations}

We used MeerKAT in UHF (580\,MHz--1.1\,GHz) with 8-s time integration. Calibration was provided by the Science Data Processor pipeline at the South African Radio Astronomy Observatory. Dispersion smearing is just 0.5\,s, less than the integration interval. We used \textsc{WSClean} to image the calibrated measurement set, selecting only baselines with lengths $>$573$\lambda$, so as to down-weight contaminating Galactic diffuse emission on scales of $>$0$\fdg1$. We imaged 9\,sq.\,deg. with a pixel scale of 1\farcs8, at a ``Briggs'' robust weighting of 0.0 \citep{1995AAS...18711202B}, outputting 10~equally-spaced channels and joining cleaning across them to account for spectral variations across the wide bandwidth due to intrinsic source spectra and the primary beam. We \textsc{clean}ed down to $3\times$ the RMS of the residuals, and then down to $1\times$ the RMS for pixels within regions already selected as containing \textsc{clean} components (\texttt{-auto-mask}$=3$; \texttt{-auto-threshold}$=1$). We imaged each 8-s correlator dump with the continuum sources subtracted and detected no single pulses to a 3$\sigma$ limit of 3\,mJy\,beam$^{-1}$. Splitting the data into 32-s intervals and folding at $P=1091$\,s, no pulsed emission was observed, to a 3$\sigma$ limit of 300\,$\mu$Jy\,beam$^{-1}$ per phase bin. The RMS of the time-integrated image is 28\,$\mu$Jy\,beam$^{-1}$. Therefore, we obtained a 3$\sigma$ upper limit on a persistent radio source of 84\,$\mu$Jy\,beam$^{-1}$.

\subsubsection{ATCA observations}

We observed with ATCA simultaneously in the C- (4--6\,GHz) and X- (8--10\,GHz) bands using the Compact Array Broadband Backend \citep{2011MNRAS.416..832W}. We applied primary (bandpass and absolute flux density scale) calibration solutions derived from PKS\,1934-638 and secondary (gain) calibration solutions from PKS\,1646-50. Imaging was performed in \textsc{Miriad} \citep{1995ASPC...77..433S}. The RMS in the C- and X-band images was 24 and 20\,$\mu$Jy\,beam$^{-1}$, leading to 3$\sigma$ upper limits of 72 and 60\,$\mu$Jy\,beam$^{-1}$, respectively. Folding the data resulted in a relatively poor 3$\sigma$ upper limit of 5\,mJy\,beam$^{-1}$ due to the imperfect $(u,v)$-coverage.
%----------------------------------------------------
\begin{figure}
\centering
\includegraphics[width =0.5\textwidth]{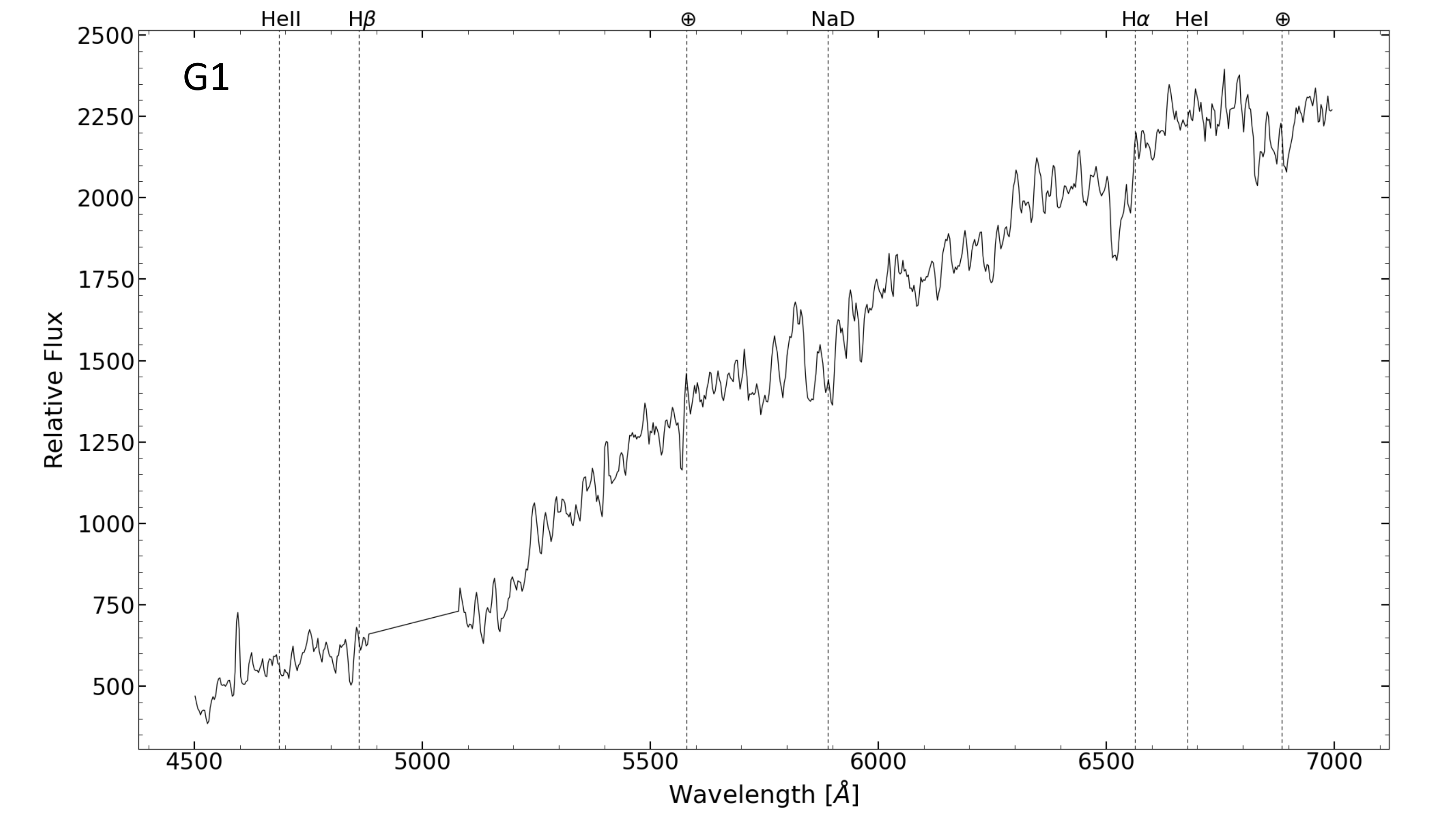}
\includegraphics[width =0.5\textwidth]{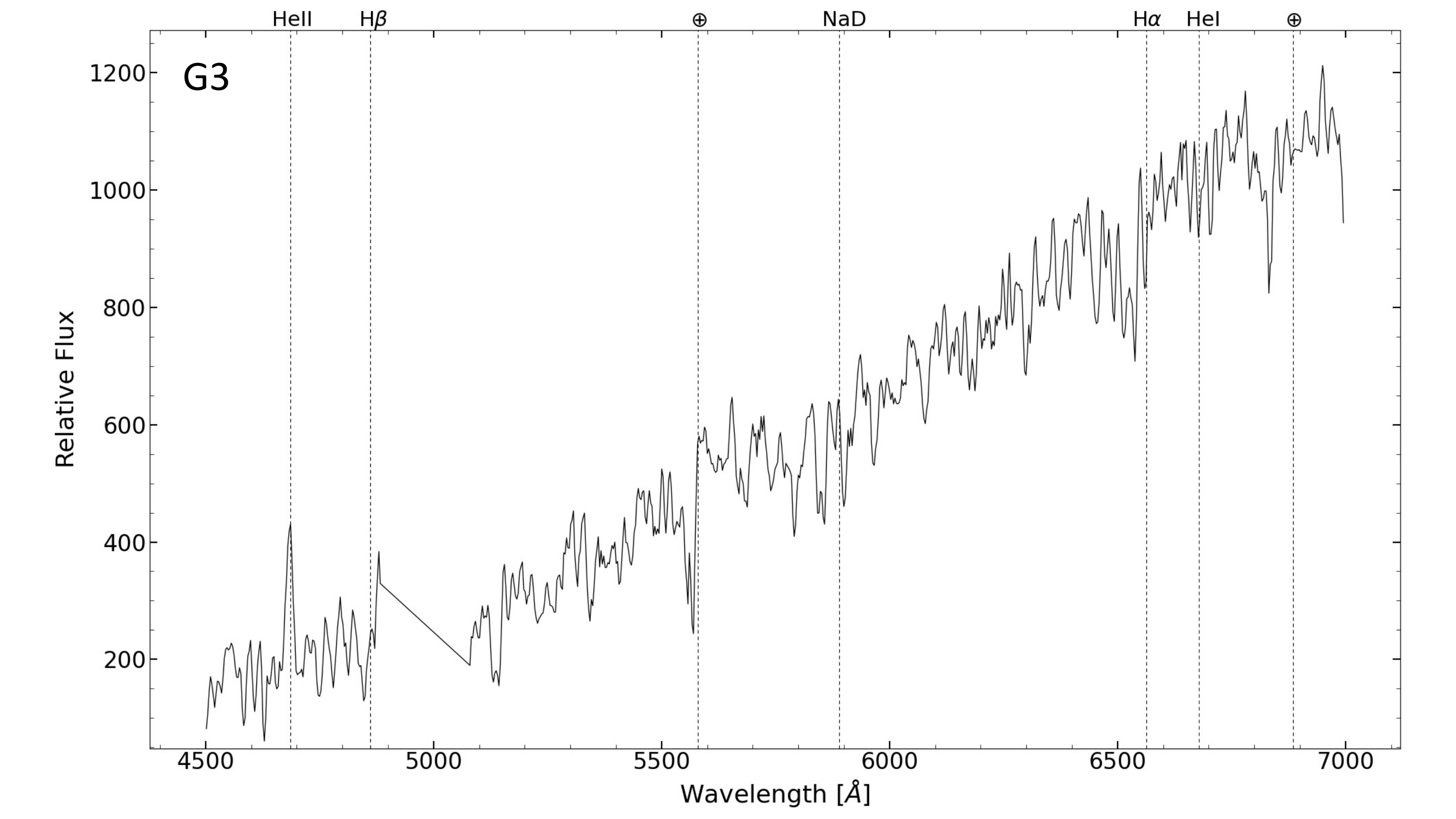}
\includegraphics[width =0.5\textwidth]{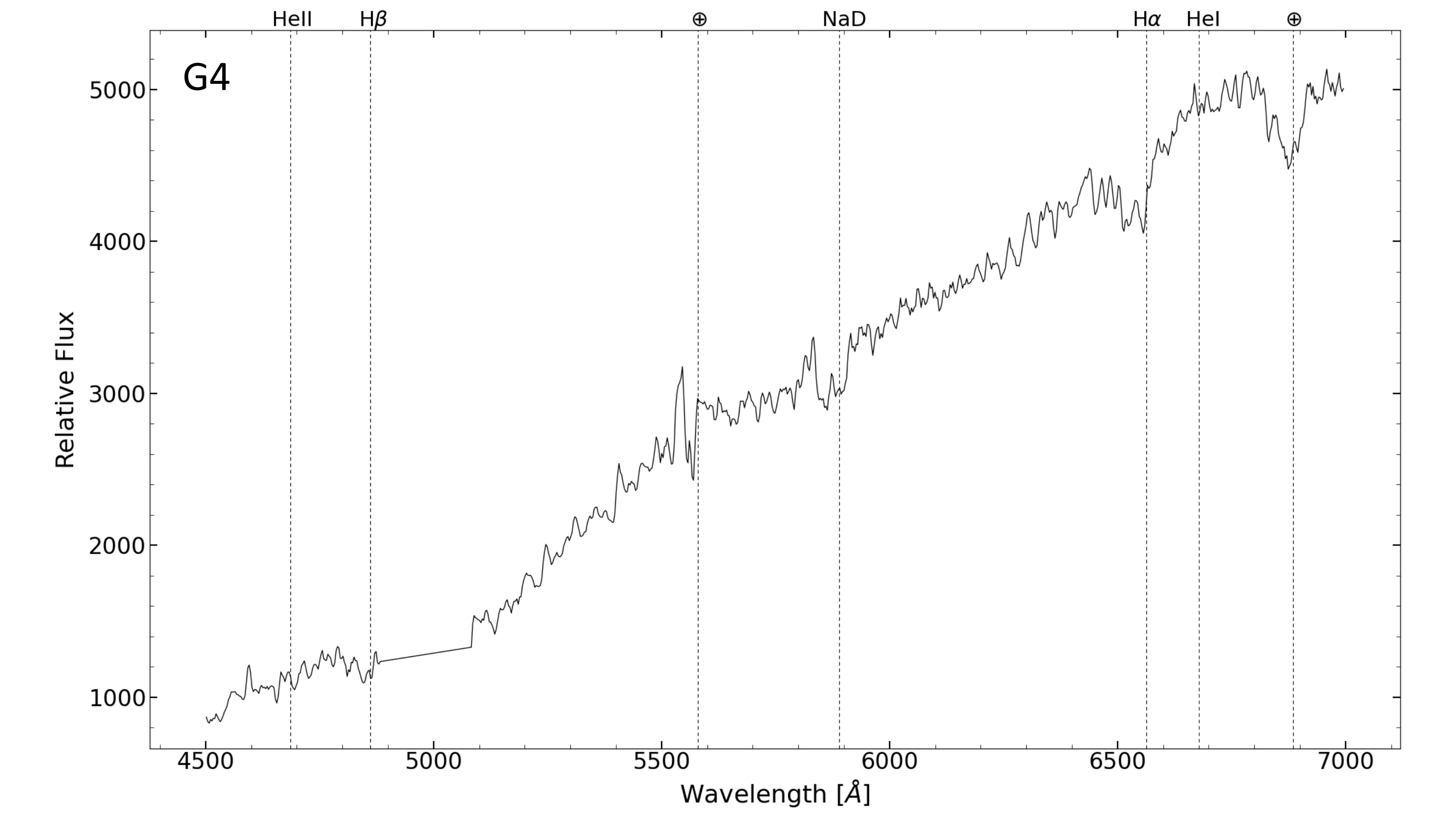}
%\vspace{-0.7cm}
\caption{{From top to bottom:} The combined average SALT spectra of G1, G3 and G4 in the range 4500-7000 \AA\ showing absorption features typical of mid-late type stars. The vertical lines report expected hydrogen Balmer (H$\alpha$ and H$_{\beta}$) and helium lines (HeI and HeII). Telluric features are also marked using an encircled cross. The gap around 5000\,\AA\ is due to the mosaicked chips of the RSS detector. A residual artifact is present in the spectrum of G3 around HeII line.}
\label{fig:salt_g134}
\end{figure}  
%----------------------------------------------------

%----------------------------------------------------
\begin{table*}[t]
\scriptsize
\centering 
\caption{Magnitudes of the five {\em Gaia} sources and the two faint optical sources within 4$\arc$ from the position of \gleam\ as derived from the VPHAS$+$, DECAPS and VVVX public surveys, as well as from recent observations at SALT. Magnitudes are in the Vega system. }
\label{tab:optstars}
\begin{tabular}{@{}cccccccccccc}
\hline
\hline
%\hline
\multicolumn{12}{c}{{\it Gaia}} \\
\hline
Source & R.A. (J2015.5) & Dec. (J2015.5)  & $d_{\rm GLEAM}$ (\arcsec) & \multicolumn{2}{c}{$m_G$} & \multicolumn{2}{c}{$Bp$} &  \multicolumn{2}{c}{$Rp$} & & \\
 \hline
G1  & 16$^\mathrm{h}$27$^\mathrm{m}$59$\fs$434 & --52$^{\circ}$35$^{\prime}$03$\farcs$58  &  0.9   & \multicolumn{2}{c}{20.24} & \multicolumn{2}{c}{21.22} & \multicolumn{2}{c}{19.37}   \\
G2  & 16$^\mathrm{h}$27$^\mathrm{m}$59$\fs$579 &  --52$^{\circ}$35$^{\prime}$06$\farcs$96    &  2.8 & \multicolumn{2}{c}{20.60} & \multicolumn{2}{c}{21.72} & \multicolumn{2}{c}{19.43}   \\
G3 & 16$^\mathrm{h}$27$^\mathrm{m}$59$\fs$870 & --52$^{\circ}$35$^{\prime}$04$\farcs$98 &     3.4   &  \multicolumn{2}{c}{20.31} & \multicolumn{2}{c}{22.02} & \multicolumn{2}{c}{19.34}   \\
G4 & 16$^\mathrm{h}$27$^\mathrm{m}$59$\fs$097 &  --52$^{\circ}$35$^{\prime}$02$\farcs$71   &   4.0  &  \multicolumn{2}{c}{18.81} & \multicolumn{2}{c}{20.04} & \multicolumn{2}{c}{17.84}  \\
G5 & 16$^\mathrm{h}$27$^\mathrm{m}$59$\fs$891 & --52$^{\circ}$35$^{\prime}$02$\farcs$44  &    4.0    &  \multicolumn{2}{c}{19.52} & \multicolumn{2}{c}{20.57} & \multicolumn{2}{c}{18.48} \\
\hline
\hline
\multicolumn{12}{c}{VPHAS$+$ survey} \\
\hline
Source  &  & & & $g$      & $r$   & $i$  & H$\alpha$ & & & & \\
\hline
G1     &  & & & 21.98(11) &  20.30(8) & 19.61(8) & 20.23(14) & & & & \\
G2     &  & & &  --       &  21.11(13) & 20.01(11) &  20.65(20) & & & & \\
G3     &   & & &  --       & 21.08(13) &  19.74(9) & 20.54(18) & & & & \\
G4     &   & & & 20.70(5) &  18.92(5)  &  17.97(3) &  18.81(5)  & & & & \\
G5     &   & & & 21.49(8) &  19.64(6)  &  18.75(4) &  19.55(8) & & & & \\
\hline
\hline
\multicolumn{12}{c}{DECAPS and VVVX surveys } \\
\hline
Source & R.A. (J2000) & Dec. (J2000)  & $d_{\rm GLEAM}$ (\arcsec)                 &  $g$        &  $r$       & $i$        & $z$        &  $Y$         &  $J$       & $H$        & $K_s$\\
\hline
G1   & 16$^\mathrm{h}$27$^\mathrm{m}$59$\fs$45 & --52$^{\circ}$35$^{\prime}$03$\farcs$5 & 0.9 &  21.731(15) & 19.906(10) & 20.226(11) & 19.250(11) & 18.256(12) & 17.733(28) & 17.088(42) & 16.784(71) \\
G2   & 16$^\mathrm{h}$27$^\mathrm{m}$59$\fs$58 & --52$^{\circ}$35$^{\prime}$07$\farcs$0 & 2.8 &  22.432(26) & 20.434(15) & 19.514(11) & 18.942(10) & 18.535(17) & 17.947(24) & 17.542(62) & 17.109(90) \\
G3   & 16$^\mathrm{h}$27$^\mathrm{m}$59$\fs$87 & --52$^{\circ}$35$^{\prime}$05$\farcs$0 & 3.4 &  22.978(45) & 20.361(14) & 19.047(8)  & 18.126(7)  & 17.744(9)  & 16.870(10) & 15.989(14) & 15.757(24) \\
G4   & 16$^\mathrm{h}$27$^\mathrm{m}$59$\fs$10 & --52$^{\circ}$35$^{\prime}$02$\farcs$7 & 4.0 &  20.567(8)  & 18.649(6)  & 17.705(6)  & 17.035(6)  & 16.750(6)  & 16.126(7) &  15.517(11) & 15.311(20) \\
G5   & 16$^\mathrm{h}$27$^\mathrm{m}$59$\fs$89 & --52$^{\circ}$35$^{\prime}$02$\farcs$4 & 4.0 &  21.221(10) & 19.297(8)  & 18.359(6)  & 17.677(6)  & 17.398(8)  & 16.764(9) &  16.185(19) & 15.937(28)\\
D1   & 16$^\mathrm{h}$27$^\mathrm{m}$59$\fs$56 & --52$^{\circ}$35$^{\prime}$05$\farcs$5 & 1.4 &     --      & 22.664(19) & 21.393(41) & 20.453(32) & 20.056(51) & 19.146(63) & 18.312(109) & 18.152(205)\\ 
D2   & 16$^\mathrm{h}$27$^\mathrm{m}$59$\fs$40 & --52$^{\circ}$35$^{\prime}$00$\farcs$9 & 3.7 &     --      & 22.625(18) & 21.671(52) & 20.777(42) & 20.500(75) &  --        &   --       &   -- \\
\hline
\hline 
%\hline 
%Source  & Proposed type & $A_V$  & $g'$       & $r'$     & $i'$    & H$\alpha$ & $J$     & $H$     & $K_s$\\
%\hline
%G1  & G8V & 2  &  22.00(11) & 20.30(8) & 19.61(8) & 20.23(14)   & 18.03(8) & 17.37(8) & 17.41(25) \\
%G2  & K5-K7 & 4 &  23.4(4) &  21.11(13) & 20.01(11) &  20.65(20) & 18.18(9) & 17.56(10) & 17.24(21) \\
%G3 & G8V-K0V & 4  & -       &  21.08(13) &  19.74(9) & 20.54(18) &  17.25(6) & 16.40(4) & 16.07(7)   \\
%G4  & F8V-G0V & 2-4  & 20.70(5) &  18.92(5)  &  17.97(3) &  18.81(5)  & 16.43(5) & 15.83(2) & 15.65(5)  \\
%G5  & K4V-K5V & 2  &     21.49(8) &  19.64(6)  &  18.75(4) &  19.55(8) & 17.09(6) &  16.55(4) & 16.39(10)  \\
\multicolumn{12}{c}{SALT} \\
\hline
 Source label    & & & &	& $r'$     		& $i'$    		& $z'$\\
\hline
G1  			& & & &	    & 20.05(3)		& 19.21(2)		& 18.64(3) & & & &  \\	
G2  			& & & &	    & 21.11(8)		& 19.54(4)		& 18.90(5) & & & &  \\
G3 			    & & & &	    & 20.32(4)		& 19.14(2)		& 18.25(2) & & & &  \\
G4  			& & & &	    & 18.75(1)		& 17.85(1)		& 17.21(1) & & & &  \\
G5  			& & & &	    & 19.40(2)		& 18.48(1)		& 17.84(2) & & & &  \\
\hline
\hline
\multicolumn{12}{c}{VPHAS$+$ 5$\sigma$ Upper limits} \\
\hline
\multicolumn{12}{c} {$u=21.6 $ \,  $g=22.8$   \,     $r=21.6$   \,   $i=20.4$     \, $H_{\alpha}=20.7$ \,  } \\
%$J=19.7$   \,   $H=19.3$   \,   $K_s=18.8$} \\
\hline
\multicolumn{12}{c}{DECAPS 5$\sigma$ and VVVX 3$\sigma$ Upper limits} \\
\hline
\multicolumn{12}{c} {$g=23.7$   \,     $r=22.9$   \,   $i=22.8$ \,  $z=22.5$ \, $Y=21.6$ \, $J=19.9$   \,   $H=19.0$   \,  $K_s=18.2$} \\
\hline
\hline
\end{tabular}

\end{table*}

\normalsize
%----------------------------------------------------

%----------------------------------------------------
\begin{figure*}
\centering
\includegraphics[width = 0.52\textwidth]{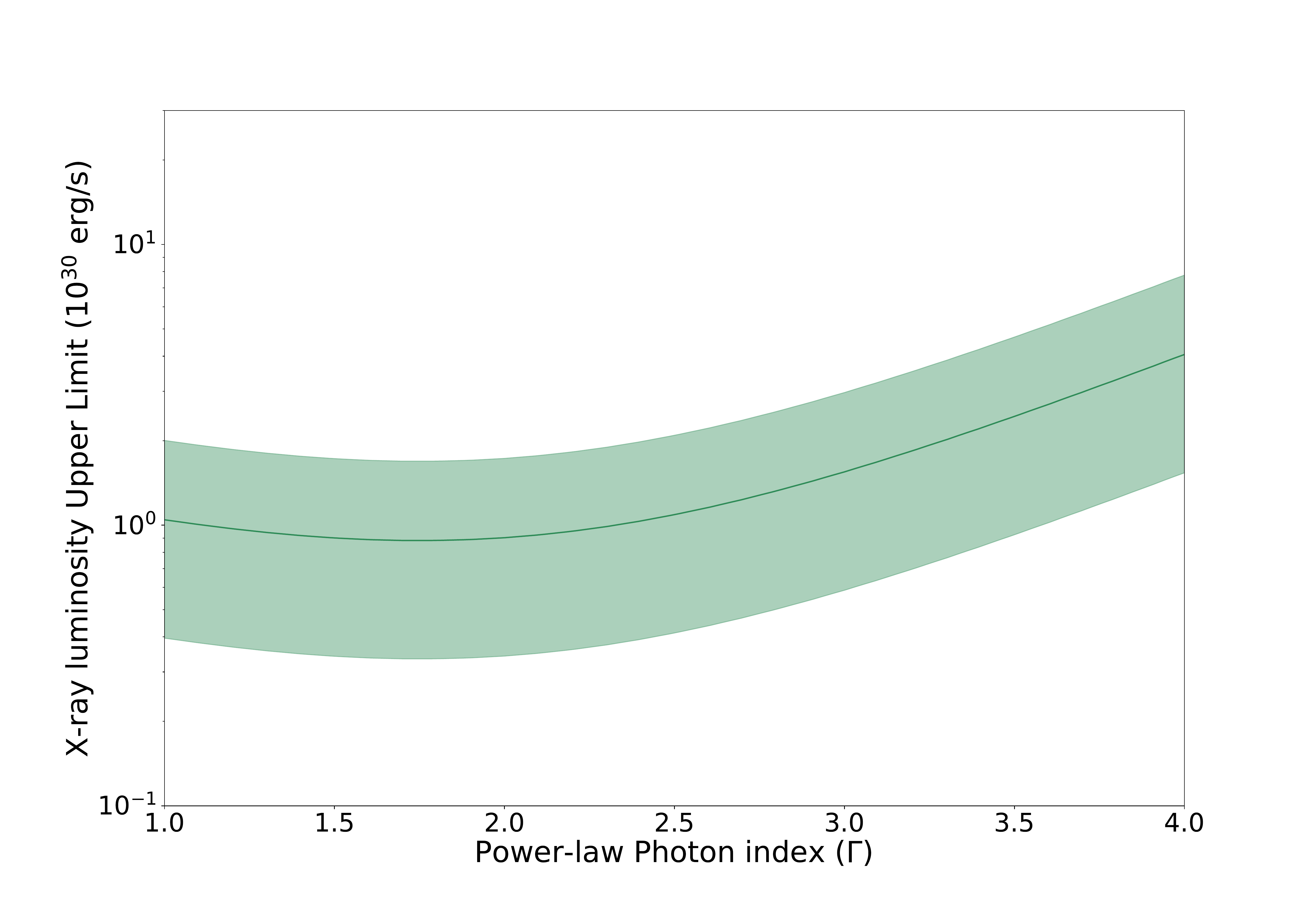}
\hspace{-1cm}
\includegraphics[width = 0.52\textwidth]{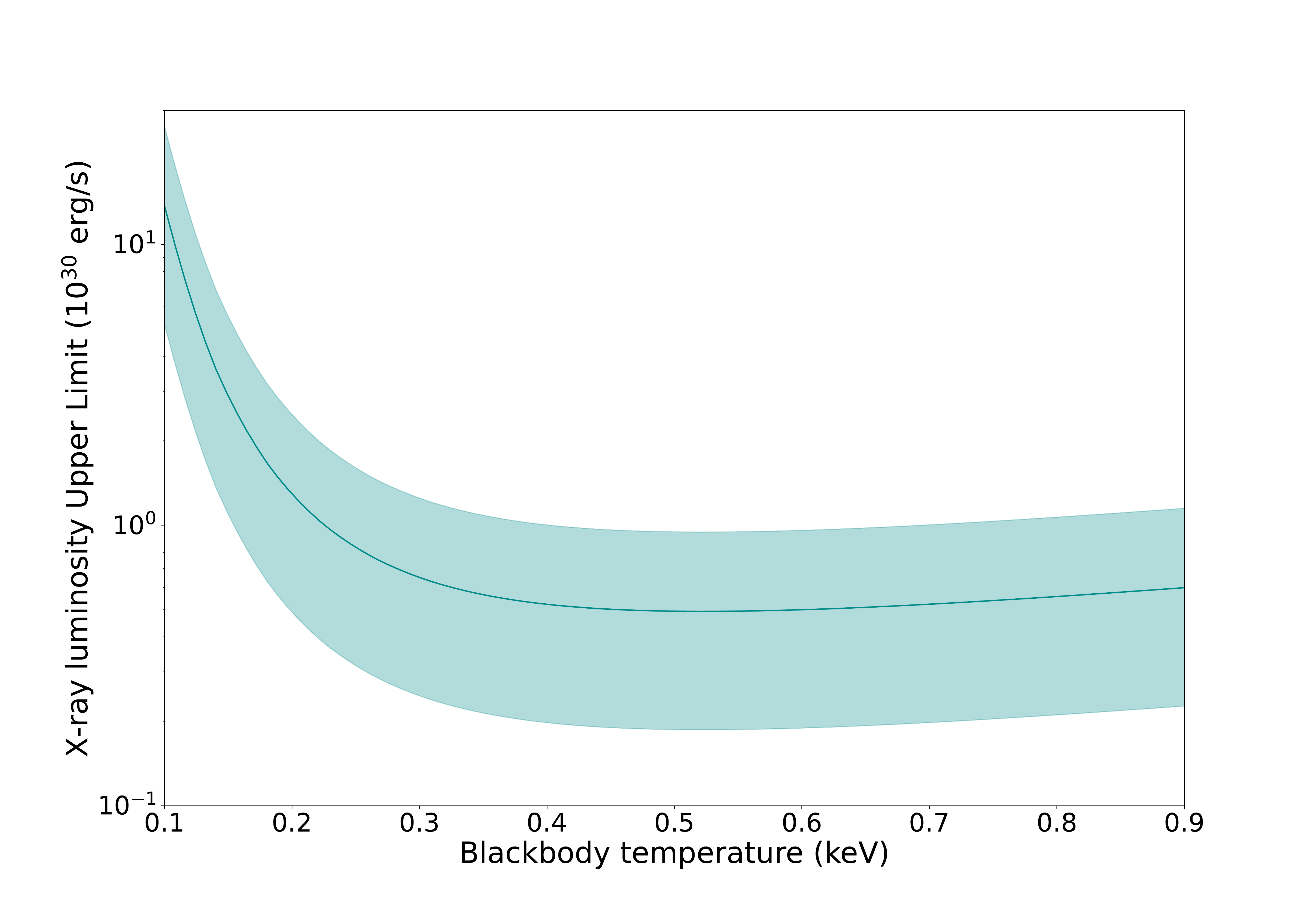}
\vspace{-0.3cm}
\caption{Upper limits at 3$\sigma$ confidence level on the X-ray luminosity of \gleam\ derived assuming either an absorbed power-law spectrum with $\Gamma=1-4$ (left panel) or an absorbed blackbody spectrum with $kT=0.1-0.9$\,keV (right panel). The solid curves mark the limits derived assuming a distance of 1.3\,kpc, while the shaded areas indicate the range of luminosities that account for the uncertainty in the source distance.}
\label{fig:X-ray_limits}
\end{figure*}  
%----------------------------------------------------

\subsection{Optical and nIR observations}
\label{sec:opt_obs}

\subsubsection{Archival imaging observations}
The field of \gleam\ has been observed in the optical band by ``The VST Photometric H$\alpha$ Survey of the Southern 
Galactic Plane and Bulge'' (VPHAS$+$; \citealt{drew14}). The public VPHAS$+$ Data Release 4 (DR4) contains several 
images of the field of \gleam\ acquired between 2016 and 2017 with the OmegaCAM imager mounted
on the 2.6-m VLT Survey Telescope (VST) in the $u$, $g$, $r$ and $i$ filters (6, 9, 12 and 6 images, respectively) 
and narrow band NB$_-$659 filter centered on H$\alpha$ (9 images).
The single exposures were 150\,s for the $u$ filter, 40\,s for the $g$ filter, 25\,s for the $r$ and $i$ filters and 
120\,s for the H$\alpha$ filter. The calibrated images and derived source catalogues were retrieved from the ESO 
archive\footnote{\url{http://archive.eso.org}}.

The field has also been covered in the optical band by ``The DECam Plane Survey'' (DECAPS; \citealt{schlafly+18}) 
with the DECAP imager mounted at the V\'ictor M. Blanco 4-m Telescope in Chile between March 2016 and May 2017 (see Fig.\,\ref{fig:decaps_FoV}). Observations were performed using five broadband filters 
$g$,$r$,$i$,$z$,$Y$. The single exposures were 96\,s for the $g$ filter and 
30\,s for the other filters, reaching much deeper limits than VPHAS$+$
(see Tab.\,\ref{tab:optstars}). The calibrated images and derived multi-band merged photometry catalogs were 
retrieved from the DECAPS archive\footnote{\url{http://decaps.skymaps.info/}}. 

The field has been covered in the nIR by ``The Vista Variables in the Via Lactea eXtended ESO 
Public Survey'' (VVVX; \citealt{minniti10}). About 200 images were acquired in the $J$, $H$ and $K_s$ filters using 
the 4-m Visible and Infrared Survey Telescope for Astronomy (VISTA) between July 2016 and September 2019, with
single exposures of 10\,s for the $J$ filter, 6\,s for the $H$ filter and 4\,s for the $K_s$ filter adopting 
different ditherings. The calibrated stacked images and derived catalogs were retrieved from the ESO archive.

We also inspected the {\em Gaia} early and final Data Releases 3 (eDR3, DR3) \citep[]{gaia21,gaia22} to search for a possible 
optical counterpart within the 2\arc\ of the radio position of \gleam\ \citep[][1-$\sigma$ confidence]{Hurley-Walker2022} and found one 
faint ($m_G$=20.24; $Bp$=21.22 $Rp$=19.37) source, here named G1. There are four additional 
bright {\em Gaia} objects nearby which we name G2, G3, G4 and G5 (see Tab.\,\ref{tab:optstars} for more details). 
Unfortunately, the parallaxes of these faint {\em Gaia} stars are undetermined,  preventing a comparison with 
the radio-derived distance of 1.3$\pm$0.5\,kpc of \gleam\, \citep{Babusiaux2022}.
The positions of \gleam, the {\em Gaia} stars and the two other
fainter stars detected in the DECAPS survey are displayed on a $z$-band DECAPS cutout image in Fig.\,\ref{fig:decaps_FoV}. In Tab.\,\ref{tab:optstars} we report the magnitudes of these stars, and to keep uniformity we have used the AB-to-Vega conversion of \cite{fukugita1996} for the $g$,$r$,$i$ and $z$ bands 
and \cite{hewett2006} for the Y-band.

We searched for optical and nIR sources close to the position of \gleam\ in the VPHAS$+$, DECAPS and VVVX catalogs.  None of the {\em Gaia} stars is detected in the VPHAS$+$ $u$ band and G2 and G3 are not detected in the VPHAS$+$ $g$ band. Tab.\,\ref{tab:optstars} reports the derived optical ($g$,$r$,$i$ and H$\alpha$) magnitudes of these
{\em Gaia} objects following the prescription for VPHAS$+$ data releases. Namely, we used the fluxes derived with an optimal aperture radius of 1\arc\ (Aperture 3), applied aperture, airmass and exposure time correction and adopted the zero points in the Vega system reported in the corresponding image catalogs. On the other hand, all the {\em Gaia} objects are detected in the DECAPS survey. This survey reaches much fainter
magnitude limits (see Tab.\,\ref {tab:optstars}), allowing us to identify two additional faint sources in the $r$, $i$, $z$ and $Y$ filters. These two sources, named D1 and D2, are located at an angular distance of 1.4\arcsec\ and 3.7\arcsec\ from \gleam, respectively (see Fig.\,\ref{fig:decaps_FoV}).
Given the large uncertainties in the VPHAS$+$ photometry, we henceforth use the DECAPS photometry to derive information on the nature of the seven detected sources. 

Furthermore, while the same procedure used for VPHAS$+$ has been adopted to the VISTA image catalogs, the two
faint sources, D1 and D2, are not detected in the single images. We then used the nIR PSF photometry 
recently performed for the VVV survey \citep{Alonso+18} and in particular for the VVVX survey (Alonso-Garc\'ia et al., in preparation) for the sky region of \gleam. The latter is obtained by stacking all images, making it possible to reach much fainter magnitudes in the $J$, $H$ and $K_s$ bands (see Tab.\,\ref{tab:optstars}).

%We then derived the optical ($g'$,$r'$,$i'$ and H$\alpha$) and 
%nIR ($J$, $H$,$K_s$)  magnitudes of the detected stars following the prescriptions for VPHAS+ and VVVX data
%releases. Namely we used the fluxes from the optimal aperture of radius 
%1\arc\ (Aperture 3),  then applied aperture, airmass and exposure time corrections and used the
%zero points in the Vega system reported in the corresponding image catalogues. The derived magnitudes are 
%reported in Tab.\,\ref{tab:optstars}.

%%%%%%%%%%%%%%%%%%%%%%%%%%%%%%%%%%%%%%%%%%%%%%%%%%%%%%%%%%%%%%%%%%%%%%%%%
\begin{figure*}
\centering
\hbox{
\includegraphics[width = 0.53\textwidth]{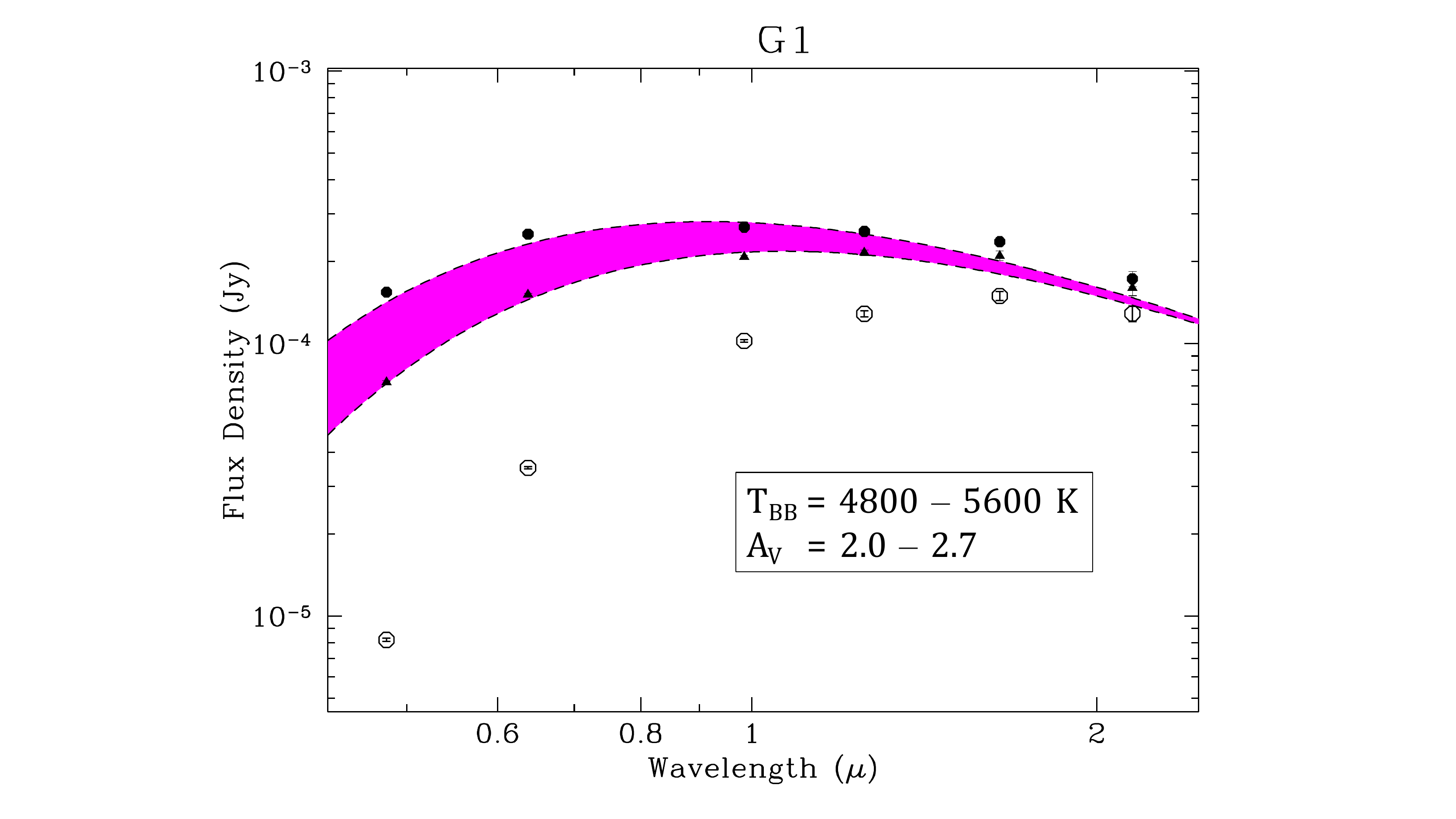}
\hspace{-1.5cm}
\includegraphics[width = 0.53\textwidth]{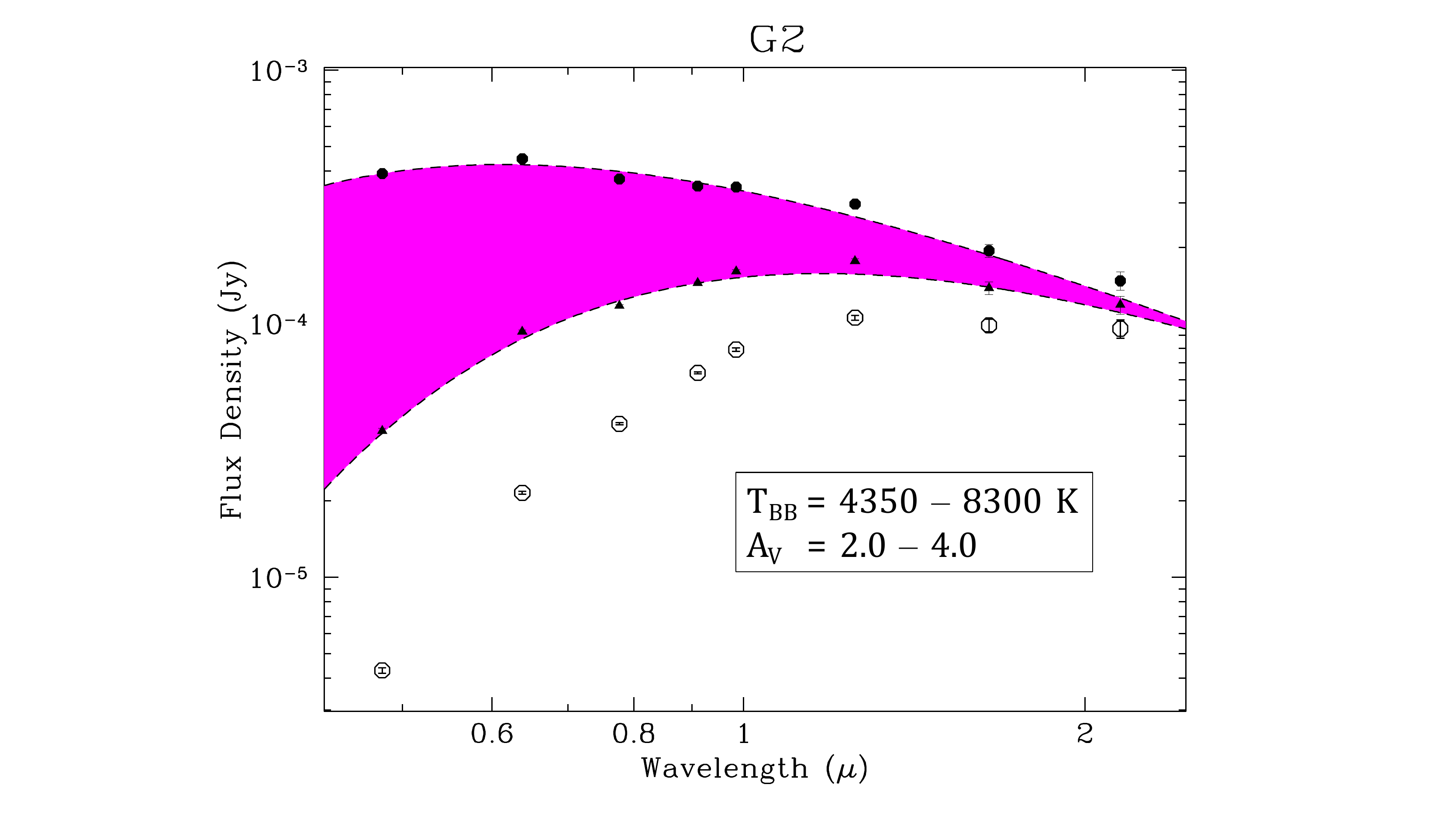}}
\hbox{
\includegraphics[width = 0.53\textwidth]{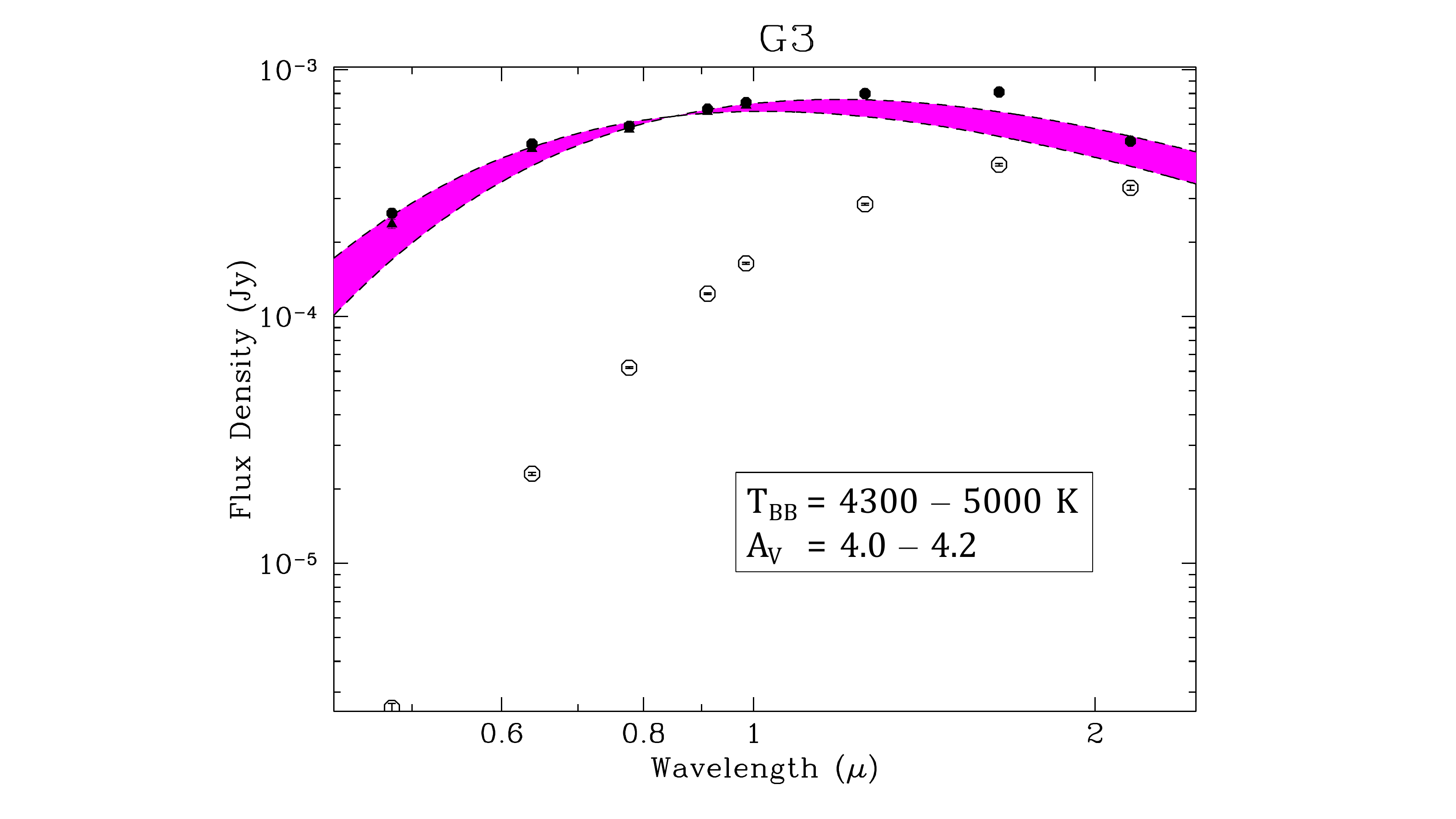}
\hspace{-1.5cm}
\includegraphics[width = 0.53\textwidth]{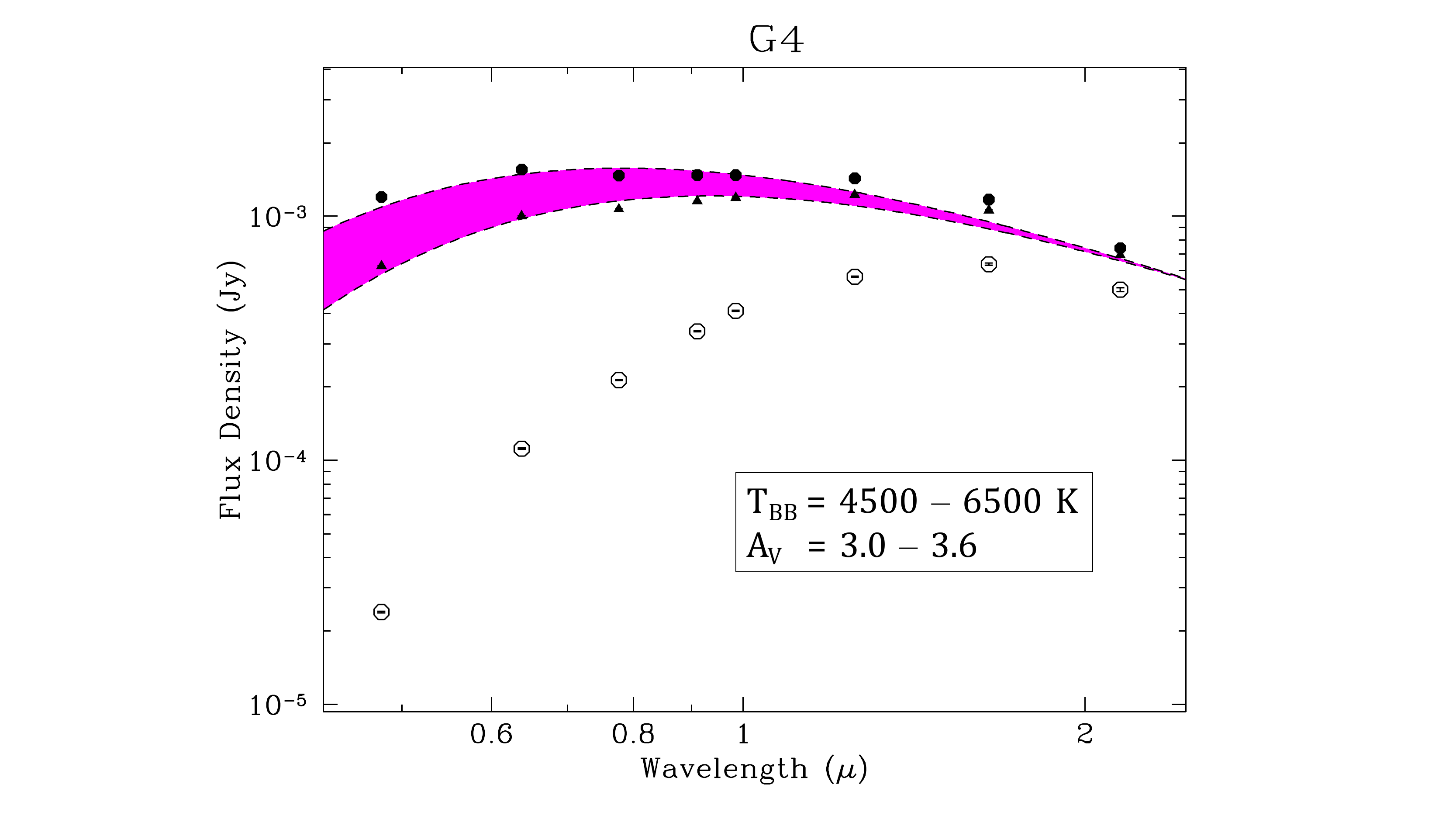}}
\hbox{
\includegraphics[width = 0.53\textwidth]{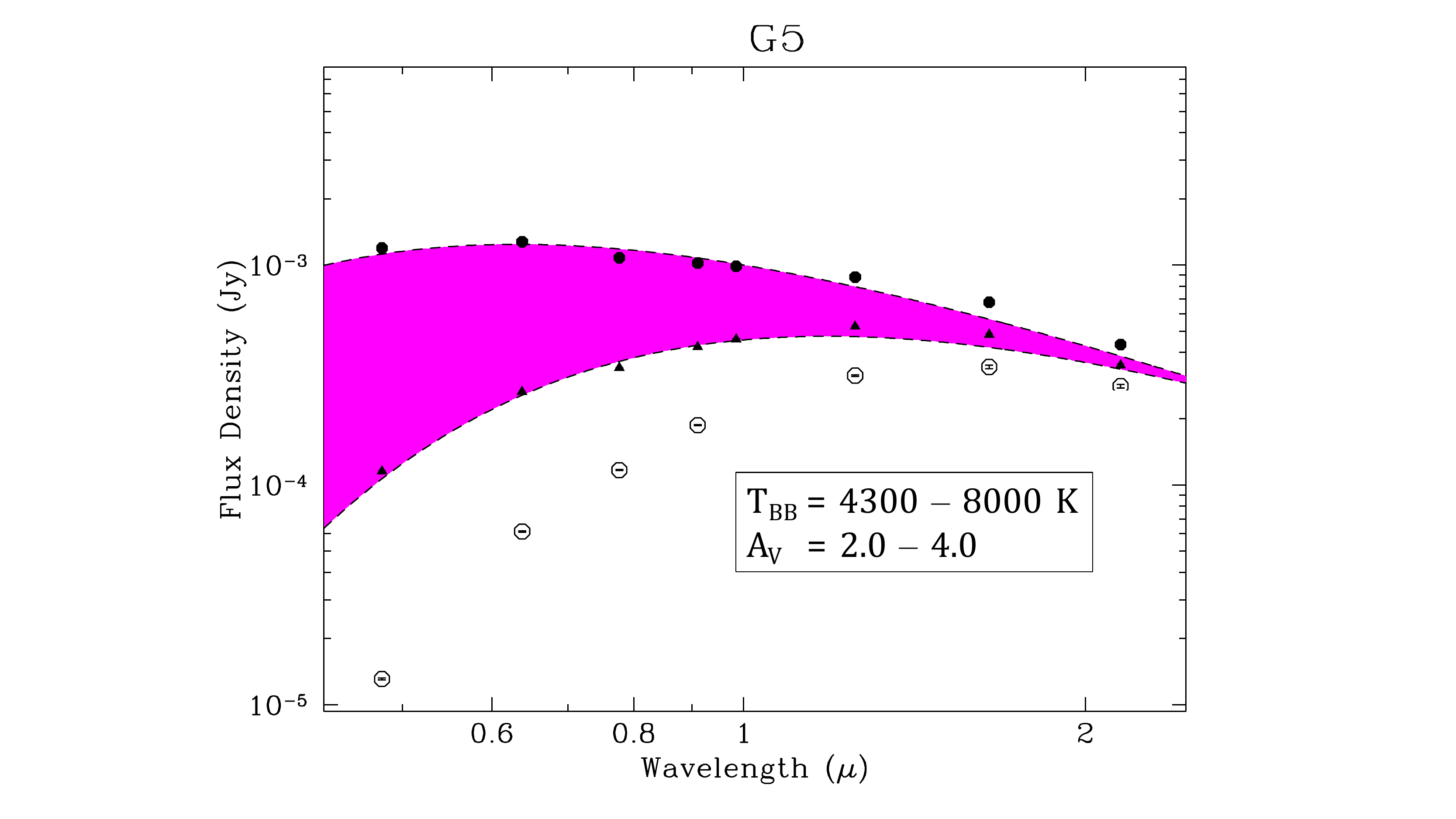}
\hspace{-1.5cm}
\includegraphics[width = 0.53\textwidth]{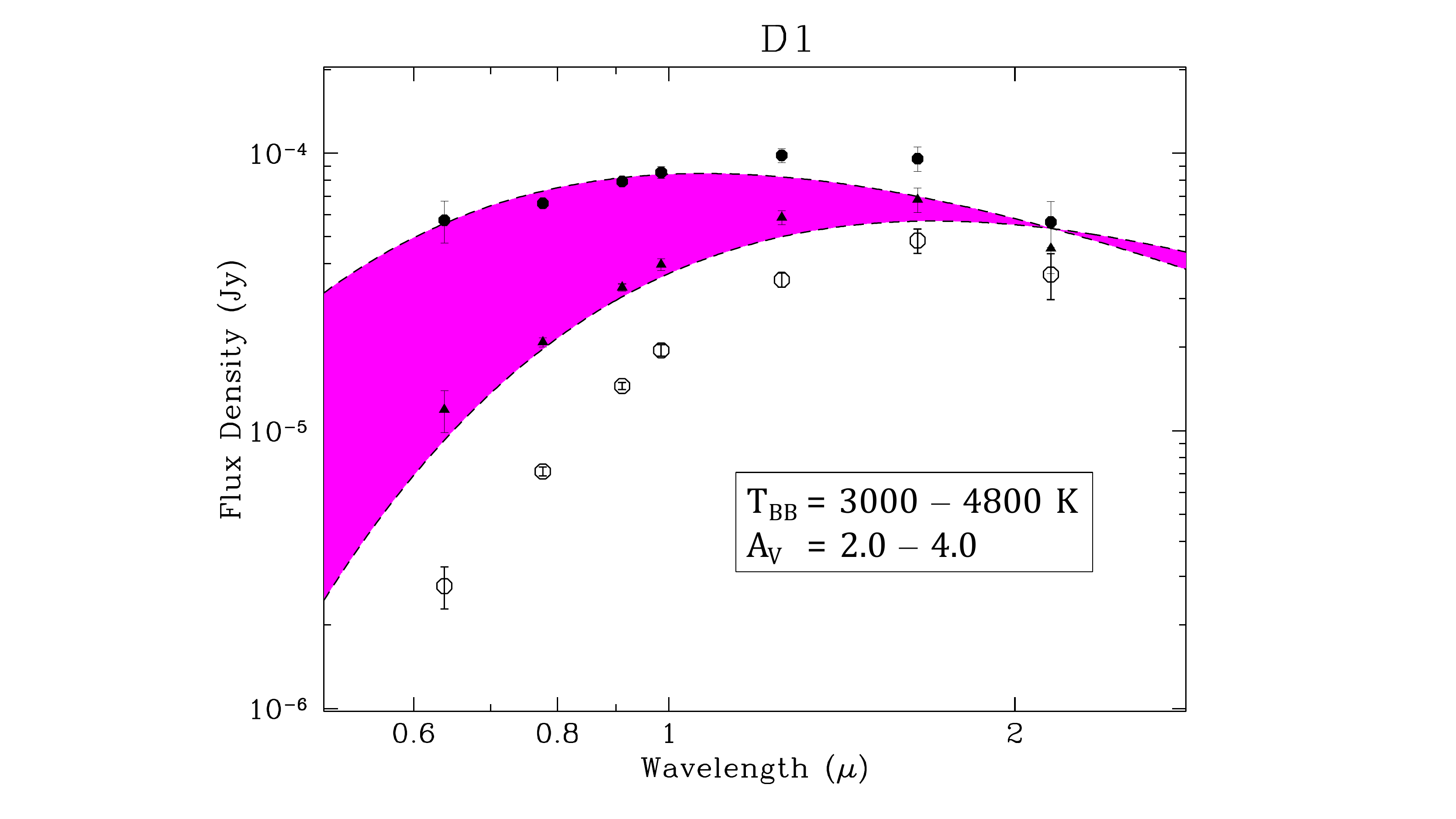}}
\hbox{
\includegraphics[width = 0.53\textwidth]{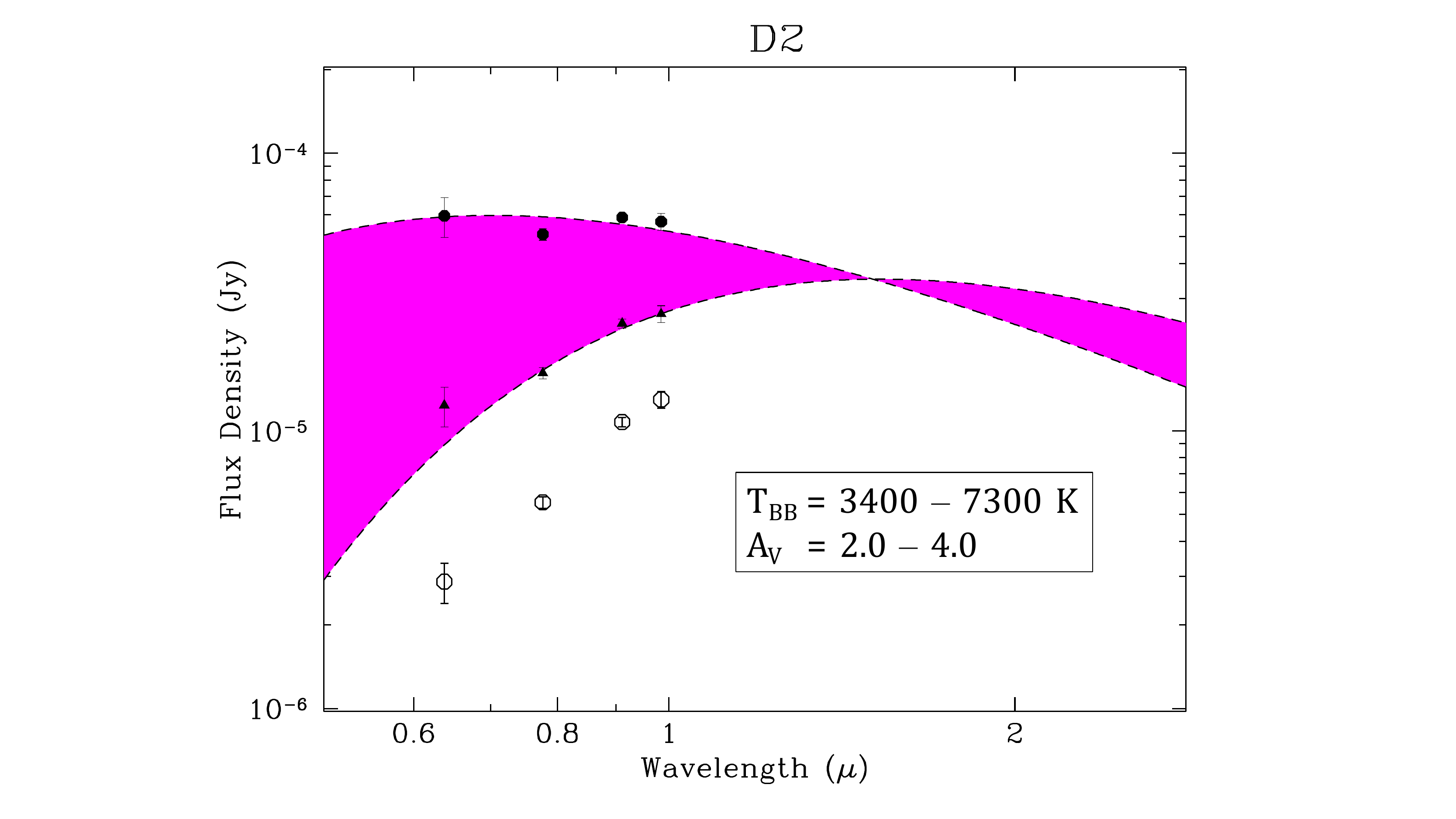}}
\caption{From left to right, top to bottom: SEDs of G1, G2, G3 , G4, G5, D1 and D2 adopting different extinction corrections, shown with
their best-fitting blackbody models. The open circles report the SEDs without reddening corrections, while the filled triangles and circles 
report the SEDs corrected for the low and high values of extinctions $A_V$, respectively, as reported in the text (see also \S\,3.2). The shaded magenta areas comprise the range of blackbody temperatures and extinctions.}
\label{fig:optsed}
\end{figure*}
%%%%%%%%%%%%%%%%%%%%%%%%%%%%%%%%%%%%%%%%%%%%%%%%%%%%%%%%%%%%%%%%%%%%%%%%%%

\subsubsection{SALT observations}

The field of \gleam\ has been observed on 2022 April 27 at the 10-m Southern African Large Telescope ({\em SALT}; \citealt{buckley06}) equipped with SALTICAM \citep{odonogue06} in $r'$ and $i'$ SDSS filters, and with the Robert Stobie Spectrograph (RSS; \citealt{burgh03}) in imaging mode with the $z'$ SDSS filter. Three images for each filter were acquired with exposure times of 120\,s ($r'$ and $i'$ filters) and 150\,s ($z'$ filter). The $r'$ and $i'$-band acquisitions were performed cyclically 
($r'$, $i'$, $r'$, $i'$, etc) from 21:37:03 to 21:56:08 UT, while $z'$-band exposures were acquired sequentially from 21:24:10 to 21:29:51 UT. A 2$\times$2 binning and Faint/Slow mode were adopted for these observations.

Images have been processed with the \texttt{Pysalt} pipeline that corrects for bias, cross-talk, gain and amplifier mosaicking. No standard stars were observed and hence no absolute flux calibration was performed due to the moving pupil of the telescope (see \citealt{buckley08}). Therefore, the images were analyzed to obtain differential photometry with the bright 2MASS stars 2MASS\,J16275876$-$5235170, 2MASS\,J16280154$-$5235061 and 2MASS\,J16275821$-$5234474 which have stable photometric measurements in DECAPS.

Aperture photometry was performed on each image and on stacked images using the three exposures in each band with the \texttt{iraf} task \texttt{daophot}. 
The targets detected in these images are the 5 {\em Gaia} stars but not the two faint ones found in DECAPS. The differential instrumental magnitudes were then converted into $r$, $i$ and $z$-filter magnitudes using the DECAPS magnitudes of the bright stars. Since the {\em Gaia} stars do not show variability within their photometric uncertainties, in Tab.\,\ref{tab:optstars} we report their magnitudes obtained from the stacked images. The associated uncertainties are statistical only. The results of the photometry of the {\em Gaia} stars appear consistent with the magnitudes obtained from DECAPS. The lack of detection of the two faint DECAPS sources is due to a shallower limiting magnitude.

Two low-resolution spectra, with a 30-minute exposure each, were acquired with the RSS mounted on {\em SALT} in long-slit mode. The first spectrum was taken starting on 2022 Feb 11 at 02:35:49 UT, while the second one starting on 2022 Feb 12 at 02:12:42 UT. The PG0300 (300\,l/mm) grating was used with a tilt angle of 5.75$^\circ$, yielding an usable wavelength range of $ \sim$3900--9000 \AA\ and a resolving power of $R\simeq$600 at 5000 \AA. We used the 1$\farcs$5 $\times$ 8\arcmin\ long-slit placed at a position angle of PA = 290$^{\circ}$ (measured from north to east) for both observations to pass through stars G1, G3, and G4 (see Fig.\,\ref{fig:salt_g134}).
The spectra were reduced using standard \texttt{iraf} tasks including flat-field, background subtraction and cosmic-ray removal. The wavelength calibration was performed using an Ar arc lamp. No standard stars were observed  and thus the spectra were not calibrated in flux.

%%%%%%%%%%%%%%%%%%%%%%%%%%%%%%%%%%%%%%%%%%%%%%%%%%%%%%%%%%%%%%%%
\section{Results} 
\label{sec:results}

\subsection{X-ray and radio}

Using the {\tt HEASARC} PIMMS tool, we estimated the 3$\sigma$  upper limit on the source X-ray flux assuming different spectral models. The expected absorption column density \nh{} from the \nh{}--DM relation by \citet{he13} is \nh{}$\simeq2\times10^{21}$\,cm$^{-2}$ using a DM = 57\,pc\,cm$^{-3}$.
%, while that expected within the Galaxy in the direction of the source is $\nh\simeq8.5\times10^{21}$\,cm$^{-2}$, \citep{willingale13}.
We derived an upper limit on the 0.3--10\,keV absorbed (unabsorbed) flux of $3.1(4.5)\times10^{-15}$\,\flux{} and $2.0(3.2)\times10^{-15}$\,\flux{} (3$\sigma$ c.l.) for a power-law with $\Gamma=2$ and a blackbody with $kT=0.3$\,keV, respectively. Assuming that the source is at a distance of 1.3\,kpc (as derived from its DM), these values translate into a luminosity limit of $9.1\times10^{29}$\,\lum{} for a power-law spectrum and $6.5\times10^{29}$\,\lum{} for a blackbody spectrum. To take into account the uncertainties related to the assumed spectral model, we make the same estimates as above but assume an absorbed power-law spectrum with a photon index ranging from $\Gamma=1-4$, as expected for rotation-powered pulsars, and an absorbed thermal spectrum modelled by a blackbody with temperature between $kT=0.1-0.9$\,keV, typical of a magnetar in quiescence \citep{rea11,cotizelati18}. In Fig.\,\ref{fig:X-ray_limits} we plot our upper limits as a function of the different assumed spectral shapes, accounting also for the uncertainty on the distance ($1.3\pm0.5$\,kpc).

The 3$\sigma$ upper limits on the radio flux density at our observed frequencies are given in the final column of Tab.\,\ref{tab:radio_obslog}, with a strong limit of $\lesssim$100\,$\mu$Jy\,beam$^{-1}$ at GHz frequencies.

%%%%%%%%%%%%%%%%%%%%%%%%%%%%%%%%%%%%%%%%%%%%%%%%%%%%%%%%%%%%%%%%

%\section{Results}

\subsection{Optical and nIR}

Fig.\,\ref{fig:salt_g134} shows the optical spectra of G1, G3 and G4.
These spectra do not reveal emission lines that could point to an interacting binary or to the
presence of deep absorption features from a hot companion star, such as a white dwarf. Note that a
sky artifact is present at HeII for the G3 star. The SALT spectra were compared with the stellar spectral library of \cite{jhc84}, which first suggest a spectral type around mid/late G for G1, a late-G or early-K star for G3 and a late F or early/mid G for G4. The low resolution of the spectra does not allow us to better
constrain the nature of these stars.
In aid of these, we have compared the observed VPHAS$+$ and DECAPS ($g$-$r$,$r$-$i$) 
colors  with the VST/OmegaCAM synthetic colors for main sequence stars tabulated in \citealt{drew14} for 
extinctions $A_V$  from 0 to 8 in steps of 2 and adopting the mean Galactic reddening law $R_V=3.1$. 
The total extinction in the direction of the source is estimated to be $A_V$ = 4.15 \citep{schlafly11}. 
However, the observed colors do not match a single sequence, indicating that these stars suffer different extinctions likely due to different distances. The same is true for the nIR ($J$-$H$,$H$-$K$) colors when compared to the main
sequence stars in the 2MASS $J$, $H$ and $K_s$ bands by \cite{straizys09} applying extinctions ranging 
from $A_V$ = 0, 2 and 4 and using the VISTA calibration from 2MASS \citep{gonzalez-fernandez18}.
Given the deeper and more accurate DECAPS photometry and the PSF VISTA nIR photometric measurements, we extracted the spectral energy distributions (SEDs) for each of the seven stars applying an extinction correction from $A_V$ 2 to 4.
%Furthermore, the large uncertainties in the VPHAS$+$ and VVVX and the lack of detection for D1 and D2 in the $g$ band and for D2 also in the nIR, do not allow us to derive constraints on the seven optical sources.
%We therefore extracted the spectral energy distributions (SEDs) using the more accurate DECAPS photometry and the VISTA nIR measurements by applying extinction corrections ranging from $A_V$ 2 to 4 for each of the seven stars. 

Fig.\,\ref{fig:optsed} reports the SEDs for these objects together with the best fitting blackbody functions for
each extinction correction. The SEDs are constructed with few measurements, giving a low quality 
of the fits ($\chi^2_{\rm red}$ ranging from 5 to 10). 
Given the unknown distance and thus reddening, and relying on the optical spectra
acquired for G1, G3 and G4, we tentatively ascribe for G1 $A_V$ between 2 and 2.7 with temperature in the range 4800--5600\,K and thus a mid G to early K spectral type; for G3  $A_V\sim$ 4 with temperature $\sim$4300\,K and thus a mid K spectral type, slightly later than estimated from spectra, or otherwise a higher reddening $A_V\sim4.2$ to
match a late G or early K type; for G4 $A_V$ should be between $\sim$3--3.8 with temperature $\sim$5500--6300\,K 
to match the late F or mid G spectral type.
For G2 we do not have spectra to constrain the spectral type and thus the extinction, $A_V$, could be between 2--4. This gives a wide range of temperatures, $\sim$4400--8300\,K, which correspond to spectral types ranging
from mid or late A to mid K. The case of G5 is similar. Therefore, for $A_V$ between 2--4, the temperature would result in the range 4300--8000\,K spanning spectral types between late A to mid K. 
Also, for D1 and D2, an extinction $A_V$ between 2 and 4 would result in blackbody temperatures in the range 
3000--4800\,K and 3400--7300\,K, respectively. The spectral types of these two faint objects, if they are main sequence stars, would range from mid M to early K and from mid M to early F, respectively.

%
%----------------------------------------------------
\begin{figure*}[t]
%\centering
%\hspace{-1cm}
%\vspace{-1.5cm}
\includegraphics[width=0.9\textwidth]{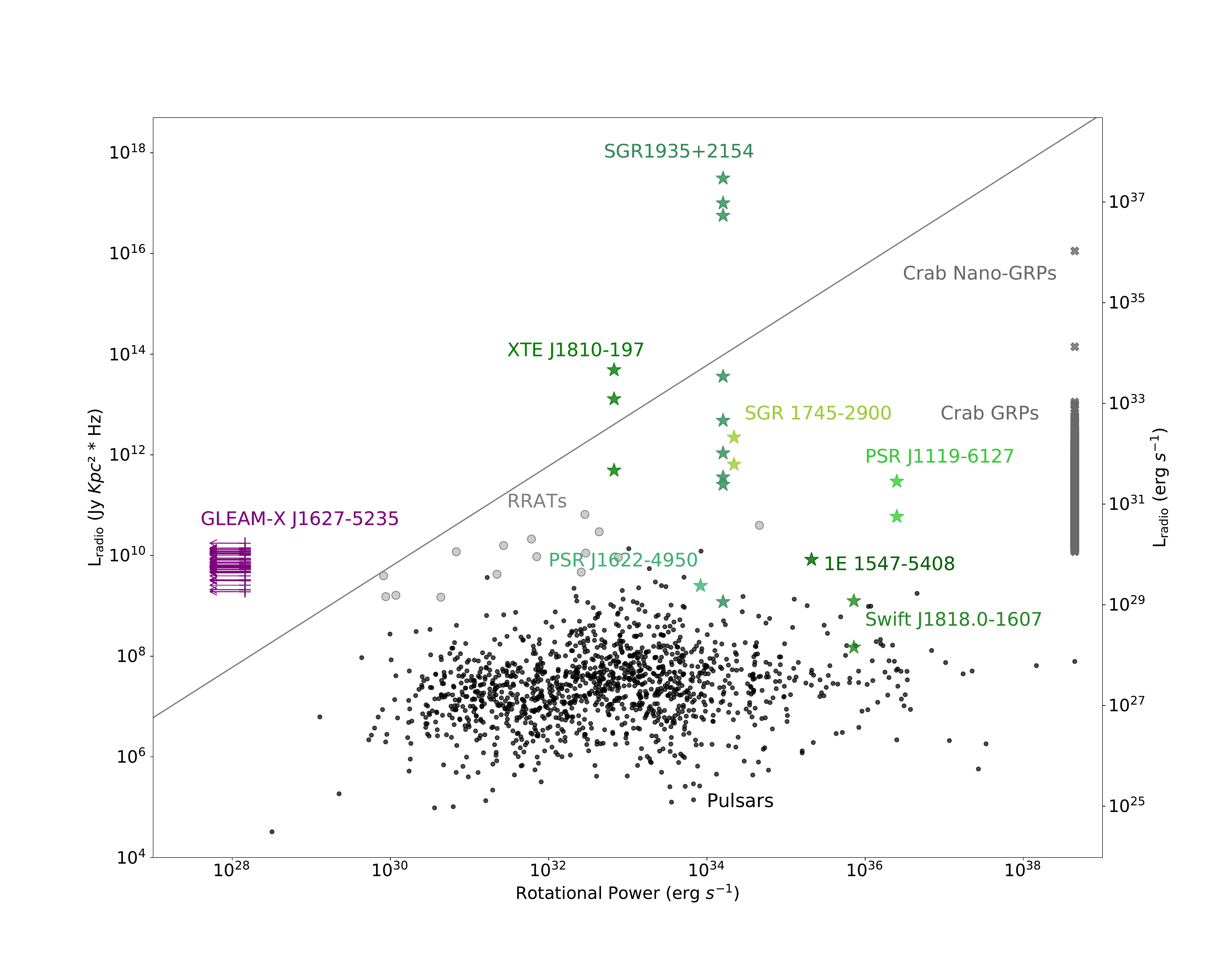}
\vspace{-1cm}
\caption{Radio luminosity versus rotational power of bright single peak emission for all pulsar classes and the \gleam's upper limits. Radio magnetars' bright single pulses are labelled in green. The gray solid line marks the relation $L_{\rm radio} = \dot{E}_{\rm rot}$.}
\label{fig:radio_edot}
\end{figure*}  
%----------------------------------------------------

%
%----------------------------------------------------
\begin{figure*}
%\centering
%\hspace{-1cm}
%\vspace{-1.5cm}
\includegraphics[width=0.9\textwidth]{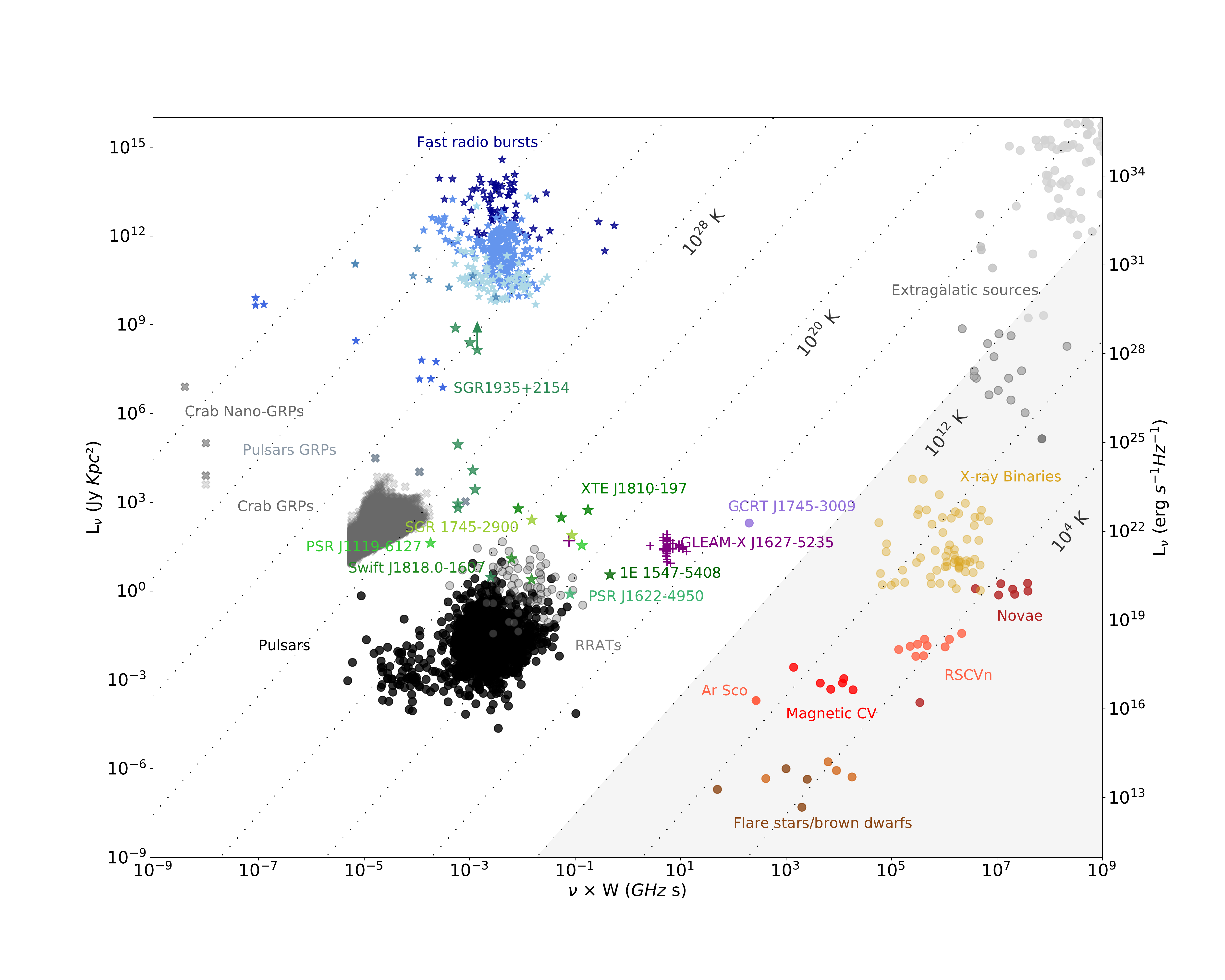}
\vspace{-1cm}
\caption{Radio-transient plane including all transient sources with a particular focus on radio magnetars' bright single pulses (different green tones), \gleam\ (violet) and white dwarf systems (different red tones). Data collected from: \cite{Camilo2006,Camilo2007,Weltevrede2011,Deller2012,Lynch2015,Majid2017,Pearlman2018,Lower2020,Esposito2021, pietka15,marsh16,Keane2018,Nimmo2022}.}
\label{fig:radio_transient_plane}
\end{figure*} 
%
%----------------------------------------------------

%%%%%%%%%%%%%%%%%%%%%%%%%%%%%%%%%%%%%%%%%%%%%%%%%%%%%%%%%%%%%%%%
\section{Discussion} 
\label{sec:discussion}

%Summary of results
In this work, we have presented simultaneous X-ray and radio observations of the 18-min radio transient \gleam, and deep optical and nIR observations of the field.

We have derived the deepest X-ray upper limits on its emission, an important ingredient in constraining its nature. The exact X-ray luminosity limit strongly depends on the assumed spectral shape and distance: we hereafter assume $L_{X} \leq 10^{30}$\,\lum{} as an average value over the different spectral models and distance errors (see Fig.\,\ref{fig:X-ray_limits} for the exact calculations). The radio limit in quiescence we have derived from the MeerKAT observations, assuming isotropic emission and a flat spectrum in the observing radio band, resulted in a quiescent radio luminosity limit of $L_{\rm radio} \leq 10^{25}$\,\lum. This radio limit is very low but not unusual in the pulsar population (see also Fig.\,\ref{fig:radio_edot}). We note that these limits are derived under strong assumptions, which may not necessarily be correct, but are only presented to give an idea of the order of magnitude.

Furthermore, the optical and nIR observations of the field revealed several objects potentially compatible with the source position (see Fig.\,\ref{fig:decaps_FoV}), that we discuss in the following sub-sections.

\subsection{The transient periodic radio emission in the framework of radio magnetars}

The observed radio characteristics of \gleam, namely its transient radio emission, its bright and variable single peaks, the pulse profile variability, and its high linear radio polarization \citep{Hurley-Walker2022}, are perfectly in line with what is typically observed for radio-loud magnetars. In fact, radio emission in magnetars is typically observed in coincidence with their X-ray outbursts, and with large variability in terms of luminosity and shape of their single peaks (see \citealt{Esposito2021} for a recent review, and references therein). The non-detection of an X-ray outburst at the time of \gleam's radio activation is not surprising, since it might have been easily missed due to the sparse and shallow coverage of the large-field-of-view X-ray monitors. The only apparent inconsistency between \gleam\ and the population of radio-loud magnetars is thus far its 18-min periodicity (see Fig.\,\ref{fig:P_Pdot_B}).

However, as extensively studied by \cite{ronchi22}, the 18-min periodicity of \gleam's radio emission cannot be reconciled with the pulsar scenario when only dipolar losses and a typical crust$+$core field configuration are considered. This would require the assumption that this pulsar has an unreasonably large magnetic field ($B\sim10^{16}-10^{17}$\,G) that does not decay in time (something unseen in the pulsar population; see also Fig.\,1 by \citealt{ronchi22}). In a typical crust$+$core field configuration, the magnetic field is expected to decay on a time scale of 10--100\,kyr depending on its intensity (the stronger the field, the faster it decays; see the spin-period evolutionary curves in Fig.\,\ref{fig:P_Pdot_B}). 

A more plausible possibility is that fall-back accretion from the supernova could easily have slowed a magnetar with a magnetic field of $\sim$10$^{13}$--10$^{14}$\,G down to its current period of 18\,min in $\sim$10$^{4}$--10$^{6}$\,yr. 
In this scenario, the supernova fossil-disk is now inactive (because the disk is now too cold or has been completely disrupted), so the source had resumed its dipolar-driven rotation and normal radio-loud magnetar activity. However, its spin period has been driven at a longer value than that of its peers at the same age and field ($\sim 10^{14}-10^{15}$\,G; for detailed simulations, see \citealt{ronchi22} and \citealt{gencali22,tong22}).

Several studies discussed the radio luminosity of \gleam\ during its radio outburst in comparison with the limits of its rotational energy \citep{Hurley-Walker2022,Erkut22}. In particular, assuming isotropic emission, the radio luminosity of the brightest single peaks ($L_{\rm radio}\sim 10^{30}-10^{31}$\,\lum; \citealt{Hurley-Walker2022}) exceeds the limits on the rotational power of the source by a few orders of magnitude. Fig.\,\ref{fig:radio_edot} shows those peak radio luminosities and the rotational power of \gleam\ in comparison with other pulsars, rotating radio transients (RRATs) and radio-loud magnetars. For the radio-loud magnetars, given their large variability, we have chosen the brightest radio pulses reported in the literature \citep[data collected from][]{Camilo2006,Camilo2007,Weltevrede2011,Deller2012,Lynch2015,Majid2017,Pearlman2018,Lower2020,Esposito2021}. It is well known that assuming isotropic radio emission is not realistic, and a beaming factor necessarily has to be present (see e.g. \citealt{Erkut22}). However, the relation between the duty cycle and the spin period of canonical pulsars has a large spread \citep{ATNFcatalog}. Moreover, it is observed that this relationship does not apply to radio-loud magnetars, which in general show larger duty cycles than what one would expect from the extrapolation of this tentative relation for radio pulsars to magnetars \citep[see i.e.][]{Camilo2006,Camilo2007}. To avoid the uncertainty of beaming models, which for magnetars are mostly unknown even theoretically, we plotted the isotropic radio luminosity for all the different pulsar classes in Fig.\,\ref{fig:radio_edot}. From this plot, at variance with canonical radio pulsars, we see how the brightest single peaks for radio-loud magnetars might exceed their rotational powers, in line with what is possibly observed for \gleam. While not resolving uncertainties related to the exact mechanism of radio emission or the beaming factor, Fig.\,\ref{fig:radio_edot} shows that, under the assumption of isotropic emission, even for magnetars the brightest single peaks exceed their rotational energy budget. Considering all the uncertainties in the assumptions used to derive the radio luminosities plotted in Fig.\,\ref{fig:radio_edot}, \gleam's radio luminosity excess over its rotational power cannot be used as an argument for or against its neutron star nature.

Furthermore, in Fig.\,\ref{fig:radio_transient_plane} we report the radio-transient plane \citep{Cordes2004,pietka15} where we include the brightest radio single peaks observed in radio-loud magnetars and compare them with \gleam\ and other classes of radio-emitting sources \citep[data collected from][]{pietka15,marsh16,Keane2018,Nimmo2022}.
On the x-axis we report the product of the transient emission duration (or pulse width) times the frequency at which it was observed, and on the y-axis the observed flux density times distance squared. The shaded grey area, starting from brightness temperatures of $T_b<10^{12}$\,K, divides the coherent and incoherent emission processes in the Rayleigh-Jeans approximation \citep{Cordes2004, pietka15}.
As reported by \cite{Hurley-Walker2022}, the radio emission of \gleam\ occupies a region on the plane compatible with being powered by a coherent process, as also observed in other radio pulsars and radio-loud magnetars.

%----------------------------------------------------
\begin{figure*}
\centering
\includegraphics[width=0.7\textwidth]{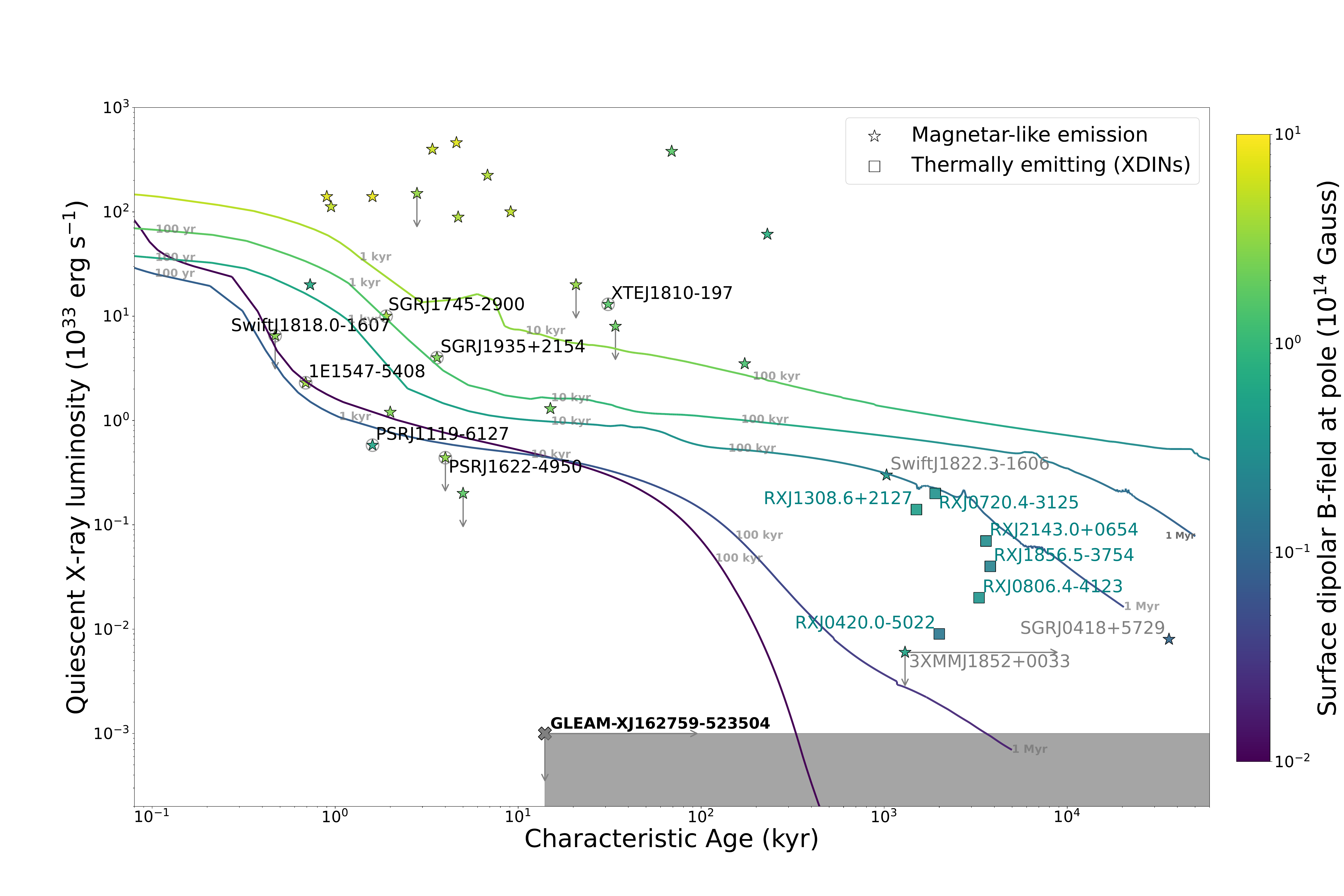}
\includegraphics[width=0.7\textwidth]{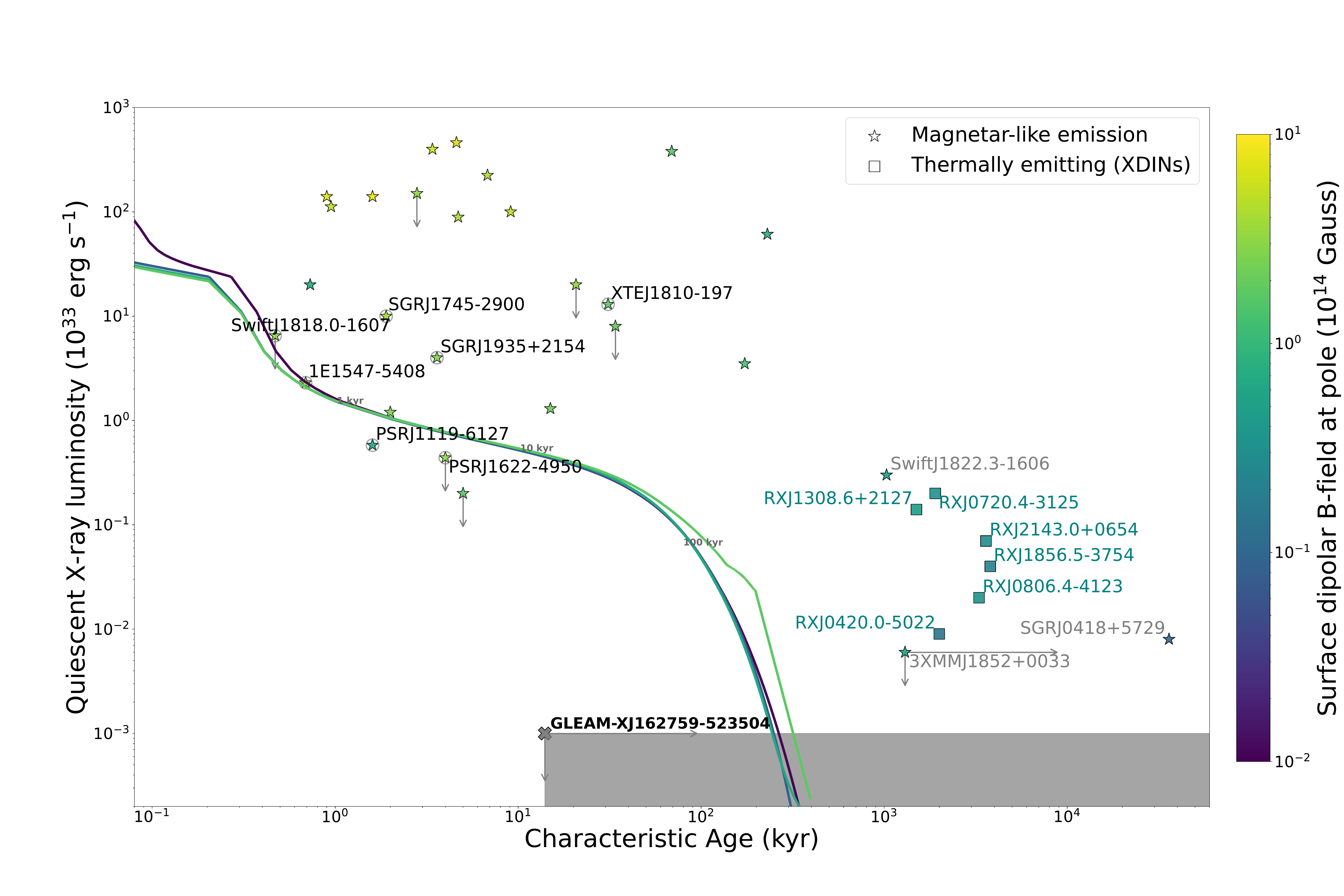}
\caption{Evolutionary tracks for crust-dominated (top panel) and core-dominated (bottom panel) $B$-field configurations superimposed on the X-ray luminosity of different pulsar classes, in particular radio-loud magnetars (labelled using a gray circle), XDINSs (in light blue) and low-field magnetars (labelled using gray names). The shaded region corresponds to the limits on the X-ray luminosity and the characteristic age derived for \gleam.}
\label{fig:cooling_modelling}
\end{figure*}  
%----------------------------------------------------

\subsection{Neutron star cooling models compared with the X-ray upper limits}

The X-ray upper limits that we derived for \gleam\ might be used to constrain its age and field configuration in the pulsar scenario. In particular, the magnetic field of a pulsar is expected to heat the neutron star crust via the dissipation of currents, which depends on the crustal microphysics, the pulsar magnetic field strength and configuration, and the star's age. To compare our X-ray upper limits with neutron star cooling curves, we used a 2D magneto-thermal evolution code \citep{aguilera2008,Pons2009,vigano2012,Vigano2013,Vigano2021}.

This 2D code assumes a background structure for the star in order to calculate necessary microphysical ingredients, such as the electron density $n_e$, thus providing realistic magneto-thermal information. A comprehensive revision of the microphysics embedded in magneto-thermal models is given by \cite{potekhin2015}. In this study, we have used the 2D magneto-thermal code to run a set of cooling models using different initial configurations. We considered: (i) crust-confined fields: the radial component of the magnetic field vanishes at the crust-core interface, while the latitudinal ($B_{\theta}$) and toroidal ($B_{\phi}$) components are different from zero; (ii) core-dominated fields: the radial component of the magnetic field is $B_r \neq 0$ at the crust-core interface, and the magnetic field lines penetrate the core. In both cases, at the stellar surface, the magnetic field is matched continuously with the potential solution of a force-free field (i.e. the electric currents do not leak into the magnetosphere). The cooling models use the Sly4\footnote{\url{https://compose.obspm.fr/}} equation of state \citep{douchin2001} with a mass of $1.4$ M$_\odot$ and the envelope model of \cite{gudmundsson1983}. The initial magnetic field ranges from $10^{12}$\,G up to $10^{15}$\,G for the dipolar poloidal component at the pole of the star, whereas the toroidal component accounts for about $40-50\%$ of the total magnetic energy in the system.  Note that we have neglected neutrino-synchrotron cooling in these simulations, as this effect requires further examination which is planned in the future.
%\textbf{Include here a sentence on how the core was actually treated considering the very different assumptions one can make (e.g., normal matter, ambipolar diffusion, superfluidity etc.).}

Fig.\,\ref{fig:cooling_modelling} shows the comparison of the 2D magneto-thermal models for crustal (top panel) and core-confined (bottom panel) field configurations, superimposed on the X-ray emission of different pulsar classes, in particular radio magnetars, normal X-ray emitting pulsars, X-ray Dim Isolated Neutron Stars (XDINSs) and low-field magnetars (data updated from \citealt{Vigano2013} to be published in Dehman et al. in prep.). In the typical scenario of a crust-confined field configuration, the X-ray limits we derive for \gleam\ are hardly compatible with any radio magnetar known so far. Even the oldest representatives of the magnetar class, the low-field magnetars, have a quiescent emission at ages $<10^{6}$\,yr that is brighter than our limits for \gleam. 
%On the other hand, if one considers field configurations confined to the neutron star core, cooling should be faster and luminosities below our limits could easily be reached.
On the other hand, we must remark that significant (observable) Joule heating effects in magnetars are only evident for crustal confined models: if the currents are mostly located in the neutron star core, Joule heating is very ineffective (most energy is lost through neutrino emission).
Hence, any of the fast cooling mechanisms discussed in the literature may be reconciled with magnetar-like fields with very low luminosities in the core-field scenario (see \citealt{anzuini22} for a similar discussion).

%----------------------------------------------------
\begin{figure*}
\centering
\includegraphics[width =9.8cm]{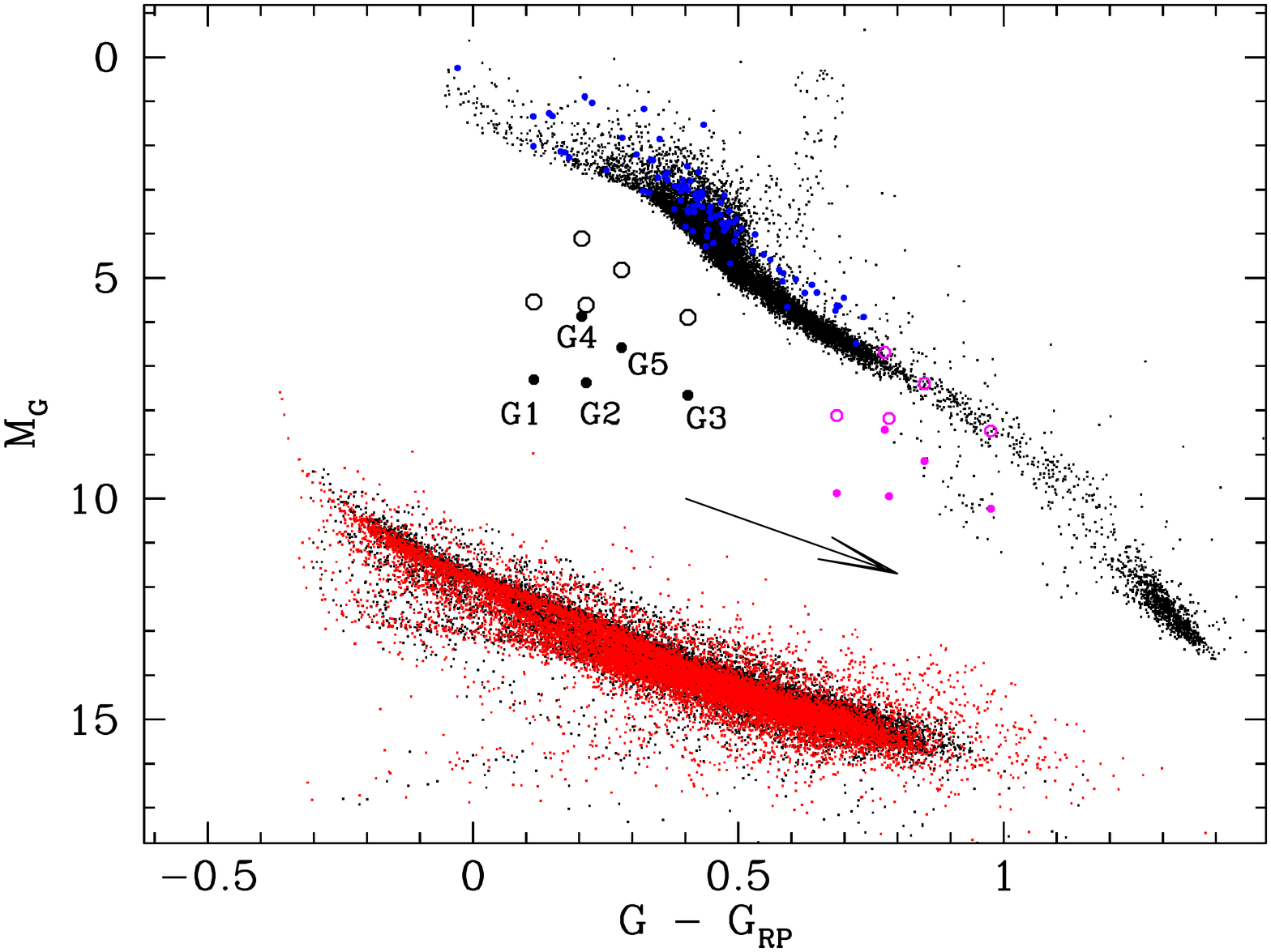}
\includegraphics[width =7.7cm]{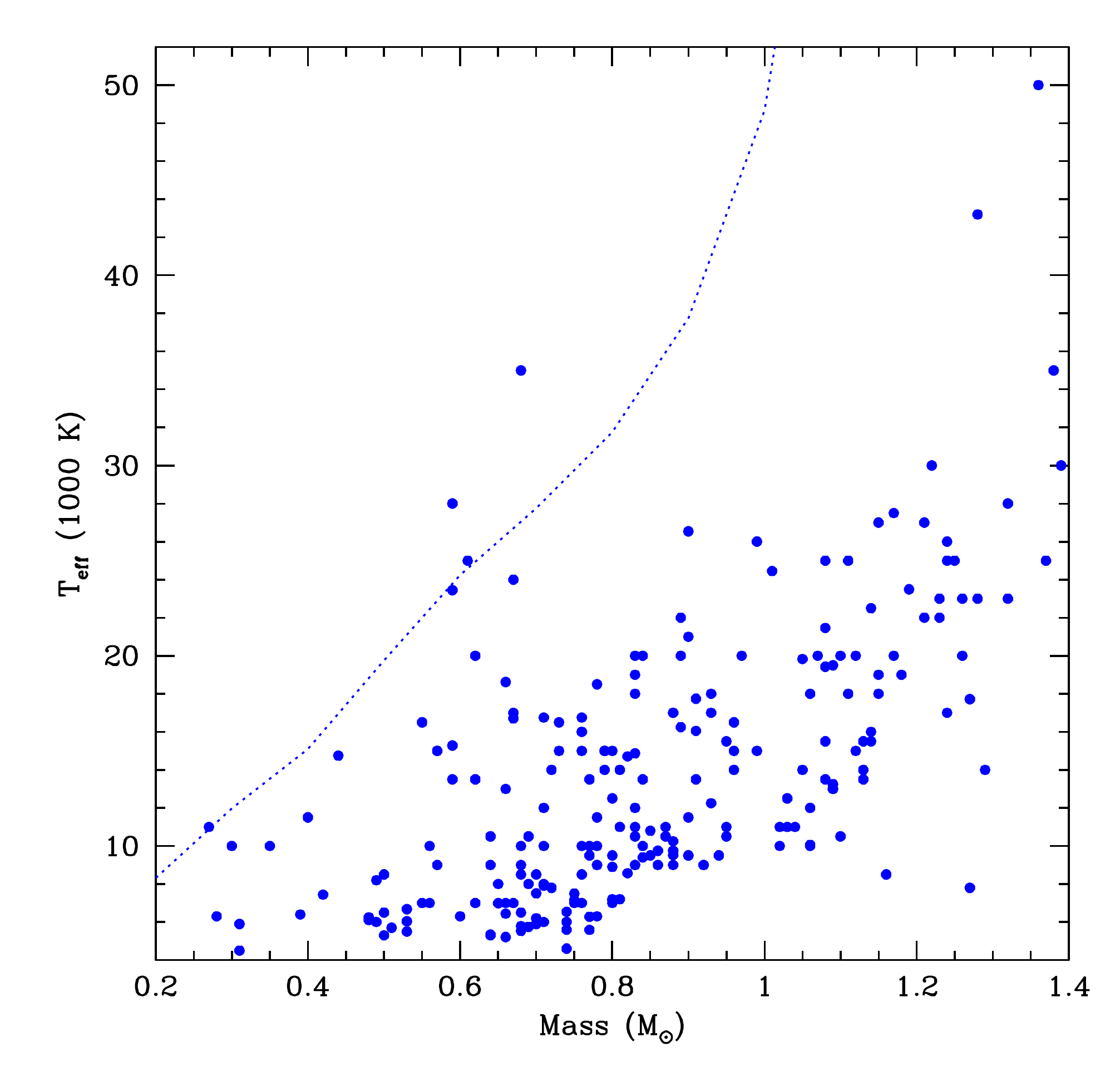}
\vspace{-0.3cm}
\caption{{\em Left panel}: Color-magnitude diagram of the five {\em Gaia} stars adopting a distance range between 0.8\,kpc (filled
circles) and 1.8\,kpc (empty circles) and extinctions $A_V =1.0$ (magenta) and $A_V = 4.1$ (black). The small dots represent the stars in the color-magnitude
diagram retrieved from the {\em Gaia} DR3 archive within 100\,pc and negligible extinction ($A_G < 0.04$). Those flagged as non-single stars are shown in blue.
The white dwarfs within 100\,pc from the {\em Gaia} DR2 catalogue of \cite{torres19} are reported as black dots and the white dwarfs within
100\,pc from the {\em Gaia} eDR3 catalogue with a probability $>90$\% of being a white dwarf \citep{GentileFusillo2021} are reported as red dots.
An extinction vector $A_V = 2$ is also reported. {\em Right panel}: Mass and temperature for known magnetic isolated white dwarfs from \cite{Ferrario2020} compared to the upper limit of $r>22.9$ at 5$\sigma$ obtained by DECam (blue dotted line).}
\label{fig:gaia_CMD}
\end{figure*}
%----------------------------------------------------

\subsection{Optical/nIR constraints and the possible white dwarf nature of \gleam }

The optical and nIR observations we report here might provide information on 
the possibility that \gleam\, is a binary system harbouring a slowly spinning magnetic white dwarf, given the detected 
radio periodicity. Radio periodic emission has so far only been observed from the white dwarf pulsar binary 
AR\,Sco \citep[]{marsh16}, although AR\,Sco is not known to display powerful transient radio pulses and this emission is clearly not coherent (unlike \gleam; see also Fig.\,\ref{fig:radio_transient_plane}).

Within the 2$\arc$ positional uncertainty, there are two optical sources, G1 and D1. Within 4$\arc$ there are five other objects. The {\em Gaia} parallaxes for the brighter five stars are undetermined, precluding a comparison with the radio distance of \gleam. Adopting an extinction $A_V$ in the range between 1.0 and 4.1 and a distance estimate of 1.3$\pm$0.5\,kpc, the positions of the {\em Gaia} stars in the
{\em Gaia} HR diagram do not match the white dwarf sequence, but fall on the main sequence of mid-late type stars for $A_V = $ 1 and 2, or below it for $A_V$ = 4 (see also the left panel of Fig.\,\ref{fig:gaia_CMD}). 
Indeed, the optical spectra acquired at {\em SALT} for G1, G3 and G4, the optical colors and the absence of detection in the $u'$ band, suggest that these optical sources are mid-late-type stars (see the left panel of Fig.\,\ref{fig:gaia_CMD}, where the positions of the five {\em Gaia} objects are shown in black for $A_V$ = 4.1 and in magenta for $A_V$ = 1). Assuming their tentative spectral types and taking into account the uncertainty in the extinction, we can place lower limits on the distances of these {\em Gaia} stars adopting the corresponding $V$-band absolute magnitudes and the {\em Gaia} DR3 conversion formulas\footnote{See Tab.\,5.8 of the {\em Gaia} DR3 data guide (\url{https://gea.esac.esa.int/archive/documentation/GDR3/}).}: for G1 $d>$3.7\,kpc, for G2 $d>$3.9\,kpc, for G3 $d\geq$1.6\,kpc, for G4 and G5 $d>$2.2\,kpc. We also note that the distances obtained for these {\em Gaia} sources by \cite{bailer_jones21} using a Galactic prior model give lower limits of 3-4\,kpc. We therefore conclude that these stars are unlikely to be emitting white dwarfs, and that only
source G3 and possibly sources G4 and G5 might have distances compatible with \gleam. Similar conclusions were drawn using the {\em Gaia} proper motions and constructing a reduced proper motion diagram given the large uncertainties in extinctions, expected absolute magnitudes and proper motions. However, this does not rule out the possibility of these sources being mid-late type companion stars of a binary white dwarf system similar to AR\,Sco. Given our nIR limits, we can only rule out such a system for $A_V \leq 2$.

The distances of the faint objects D1 and D2 cannot be constrained either. Their SEDs indicate that they are cool objects with temperatures between 3000 and 7300\,K (see Fig.\,\ref{fig:optsed}). If they were main sequence stars, their distances would be $>$3.3\,kpc for D1 and $>$4.4\,kpc for D2. Therefore, we can reliably exclude that they are companion stars of a white dwarf binary system. Similarly, the hypothesis that \gleam\ could be a hot and bright magnetic sub-dwarf star located at 1.3\,kpc \citep{loeb22} would be difficult to reconcile with the lack of detection in the DECAPS survey. We also note that the hot subdwarf stars (sdO/B type) identified by {\em Gaia} within 1.5\,kpc have absolute magnitudes between -1 and 7 in the $G$-band \citep{geier19} and are typically not highly magnetic. At the estimated distance of \gleam, such a star would have been detected.

The possibility that \gleam\ may be an isolated magnetic white dwarf pulsar located at 1.3\,kpc is less constrained. Such a star would only be detected in the optical/nIR observations if it is hot (with a temperature $>$10000\,K) and has a low extinction of $A_V \sim$ 2. The right panel of Fig.\,\ref{fig:gaia_CMD} shows the masses and temperatures of detected isolated magnetic white dwarfs with our DECAPS limit ($r>22.9$) superimposed as a dotted line \citep{Vennes2011,Kawka2012,Ferrario2020}. Therefore, we cannot rule out a scenario of an isolated, cooler white dwarf.

\section{Conclusions}
\label{sec:conclusions}

In the magnetar scenario, the upper limits we have derived on the X-ray luminosity ($L_X<10^{30}$\,\lum) imply that \gleam's age should be $>$1\,Myr for any reasonable crustal magnetic field ($B>10^{13}$\,G). The 18-min spin period (assuming a fast rotating pulsar at birth) would necessarily require a strong magnetic field and a phase of fossil-disk accretion (see \citealt{ronchi22}), but in any case the age of \gleam\ is constrained to be two orders of magnitude higher than that of typical radio-loud magnetars (which have ages $<$20\,kyr). At this age, the bright radio bursts emitted by \gleam\ would be unusual for such an old magnetar. However, it is important to note that the radio emission of \gleam\ is instead in line with what has been observed for the known population of radio-loud magnetars (which are younger). The excess of its radio luminosity over the rotational power limits is not unprecedented in the bright single pulses from radio magnetars, despite the unavoidable uncertainty due to the radio beaming factor (see Fig.\,\ref{fig:radio_edot}). We also note that, if the magnetar has a core-dominated magnetic field or has witnessed unusual fast cooling (both effects have never been unambiguously observed in a pulsar or a magnetar), the observed X-ray upper limits would be compatible with a younger age (see Fig.\,\ref{fig:cooling_modelling}). The field in the core is expected to decay on timescales of 1--10\,Myr, hence the source's spin period would initially evolve with a seemingly constant magnetic field (see Fig.\,\ref{fig:P_Pdot_B}). In this case, to explain the 18-min spin period, we would need a constant field of $B\sim10^{15}$\,G for $\sim$10$^{8}$\,yr (or $B\sim10^{16}$\,G for $\sim$10$^{6}$\,yr), which is rather extreme when compared to the known magnetar population in our Galaxy. In contrast, assuming fall-back accretion, the spin period can easily be reconciled with a 18-min value \citep{ronchi22}. Although the core-dominated field-decay interpretation or a fast cooling scenario are in principle viable, they are intriguing, as we have no evidence of other pulsars that require either hypothesis to explain their emission and spin period.

In the magnetic white dwarf scenario, our nIR and optical studies put some constraints on the binary or isolated white dwarf interpretation of \gleam. In particular, none of the sources detected within its positional uncertainty could be unambiguosly catalogued as a white dwarf or the companion star of a binary white dwarf system at 1.3\,kpc. An AR\,Sco-like system could pass unnoticed if it has a relatively large extinction. Furthermore, we exclude the possiblity of \gleam\ being a hot, bright magnetic sub-dwarf star, but cannot rule out an isolated, cooler white dwarf. We note that the rotation periods of magnetic white dwarfs are typically longer than 1\,h, with a few spinning at faster rates \citep{Kilic2021,Schwab2021}, which are believed to be the results of the merger of a double white dwarf system. Therefore a cool isolated white dwarf would need to be sufficiently magnetic, hence probably with a large mass and small radius, to produce the coherent and pulsed radio emission observed by \gleam.\\

%%%%%%%%%%%%%%%%%%%%%%%%%%%%%%%%%%%%%%%%%%%%%%%%%%%%%%
%\begin{acknowledgments}
We gratefully acknowledge the help of D. Minniti and J. Alonso-Garc\'ia in providing the VVVX nIR catalogue in advance. We thank A. Possenti for starting the literature search for some of the bright radio magnetar single pulses presented in this work, and the anonymous referee for her/his careful reading and useful comments that improved our manuscript.
NR, FCZ, CD, MR, VG, CP, AB and EP are supported by the ERC Consolidator Grant “MAGNESIA” under grant agreement No. 817661, and National Spanish grant PGC2018-095512-BI00. FCZ, AB and VG are also supported by Juan de la Cierva Fellowships. CD, MR and CA’s work has been carried out within the framework of the doctoral program in Physics of the Universitat Aut\'onoma de Barcelona. NHW is supported by an Australian Research Council Future Fellowship (project number FT190100231) funded by the Australian Government. DdM acknowledges financial support from the Italian Space Agency (ASI) and National Institute for Astrophysics (INAF) under agreements ASI-INAF I/037/12/0 and ASI-INAF n.2017-14-H.0 and from INAF ``Sostegno alla ricerca scientifica main streams dell'INAF'', Presidential Decree 43/2018 and from INAF ``SKA/CTA projects'', Presidential Decree 70/2016. DB acknowledges support from the South African National Research Foundation.
DV is supported by the ERC Starting Grant ``IMAGINE'' under grant agreement No. 948582. This work was also partially supported by the program Unidad de Excelencia Maria de Maetzu de Maeztu CEX2020-001058-M and by the PHAROS COST Action (No. CA16214).

This research has made use of the services of the ESO Science Archive Facility.
This work has made use of data from the European Space Agency (ESA) mission {\it Gaia} (\url{https://www.cosmos.esa.int/gaia}), processed by the {\it Gaia} Data Processing and Analysis Consortium (DPAC,
\url{https://www.cosmos.esa.int/web/gaia/dpac/consortium}). Funding for the DPAC has been provided by national institutions, in particular the institutions participating in the {\it Gaia} Multilateral Agreement.
This work is based on: data products from observations made with ESO Telescopes at the La Silla Paranal Observatory under programme ID 177.D-3023, as part of the VST Photometric H$\alpha$ Survey of the Southern Galactic Plane and Bulge (VPHAS$+$, \url{www.vphas.eu}); data products from VVVX Survey observations made with the VISTA telescope at the ESO Paranal Observatory under programme ID 198.B-2004. The SALT observations were obtained under the SALT Large Science Programme on transients (2018-2-LSP-001; PI: DB) which is also supported by Poland under grant no. MNiSW DIR/WK/2016/07.
The MeerKAT telescope is operated by the South African Radio Astronomy Observatory, which is a facility of the National Research Foundation, an agency of the Department of Science and Innovation. The Australia Telescope Compact Array is part of the Australia Telescope National Facility (\url{https://www.atnf.csiro.au/}) which is funded by the Australian Government for operation as a National Facility managed by CSIRO. We acknowledge the Gomeroi people as the traditional owners of the Observatory site. 
This scientific work makes use of the Murchison Radio-astronomy Observatory, operated by CSIRO. We acknowledge the Wajarri Yamatji people as the traditional owners of the Observatory site. Support for the operation of the MWA is provided by the Australian Government (NCRIS), under a contract to Curtin University administered by Astronomy Australia Limited. Establishment of the Murchison Radio-astronomy Observatory and the Pawsey Supercomputing Centre are initiatives of the Australian Government, with support from the Government of Western Australia and the Science and Industry Endowment Fund. We acknowledge the Pawsey Supercomputing Centre which is supported by the Western Australian and Australian Governments. Access to Pawsey Data Storage Services is governed by a Data Storage and Management Policy (DSMP). ASVO has received funding from the Australian Commonwealth Government through the National eResearch Collaboration Tools and Resources (NeCTAR) Project, the Australian National Data Service (ANDS), and the National Collaborative Research Infrastructure Strategy. 
%This research has made use of NASA’s Astrophysics Data System Bibliographic Services.
%\end{acknowledgments}

\facilities{\cxo\ (ACIS-S), ATCA, Blanco Telescope (DECAPS), Gaia, MeerKAT, MWA, SALT, VISTA (VVVX), VST (VPHAS$+$).} %\vspace{-0.5cm}
%\software{CIAO v.4.14 and CALDB v.4.9.6 \citep{fruscione06}; {\sc aoflagger} and {\sc cotter} \citep{2012A+A...539A..95O}; \textsc{WSClean} \citep{2014MNRAS.444..606O,2017MNRAS.471..301O}; {\sc Aegean} \citep{2018PASA...35...11H}; {\sc miriad} \citep{Miriad}; {\sc TopCat} \citep{Topcat} \textsc{NumPy} \citep{NumPy,harris2020array}; \textsc{AstroPy} \citep{Astropy}; \textsc{SciPy} \citep{SciPy}, \textsc{Matplotlib} \citep{Matplotlib}. }

%%%%%%%%%%%%%%%%%%%% REFERENCES %%%%%%%%%%%%%%%%%%

%\bibliography{bibliography}

\end{document}